%% file: Brennan_SiMs_draft.tex
\documentclass[a4paper,11pt]{article}
\pdfoutput=1 

\usepackage{jheppub} 

\usepackage{lineno}
\usepackage{caption}
\usepackage{subcaption}		
\usepackage{multirow}		
\usepackage{tikz}		
\usepackage{array}      	
\usepackage{xcolor}
\usepackage{amsmath}
\usepackage{rotating}   
\usepackage{afterpage}  

\usepackage{hyperref}

\tikzstyle arrowformat=[scale=1.5]
\tikzstyle scalar =[dashed]
\tikzstyle mom=[draw=blue]
\tikzset{
gluon/.style={decorate, draw=black, decoration={coil, amplitude = 3.5pt, segment length=4.8pt}},
photon/.style={decorate, draw=black, decoration={snake=coil, amplitude = 2pt, segment length=9pt}},
zbo/.style={decorate, draw=black, decoration={snake=coil}},
fermion/.style={draw=black, postaction={decorate}, decoration={markings, mark = at position 0.55 with {\arrow[arrowformat]{>}}}},
antifermion/.style={draw=black, postaction={decorate}, decoration={markings, mark = at position 0.55 with {\arrow[arrowformat]{<}}}}
}
\usetikzlibrary{patterns}
\usetikzlibrary{arrows}
\usetikzlibrary{decorations.pathmorphing,decorations.markings, decorations.pathreplacing, trees}

\newcommand{\MG}{M{\footnotesize AD}G{\footnotesize RAPH}5 }
\newcommand{\MGnospace}{M{\footnotesize AD}G{\footnotesize RAPH}5}
\newcommand{\PYTHIA}{P{\footnotesize YTHIA} 8 }
\newcommand{\PYTHIAnospace}{P{\footnotesize YTHIA} 8}
\newcommand{\FASTJET}{F{\footnotesize AST}J{\footnotesize ET}}
\newcommand\T{\rule{0pt}{2.6ex}}       				
\newcommand\B{\rule[-1.2ex]{0pt}{0pt}} 				
\newcommand{\met}{\ensuremath{E_{\mathrm{T}}^{\mathrm{miss}}}}		
\newcommand{\metvec}{\vec{E}_{\mathrm{T}}^{\mathrm{miss}}}	
\newcommand{\mDM}{m_{\mathrm{DM}}}
\newcommand{\mX}{m_{\chi}}
\newcommand{\Mmed}{M_{\mathrm{med}}}
\newcommand{\Mstar}{M_{\star}}					
\newcommand{\monoZ}{mono-$Z$(lep) }
\newcommand{\monojet}{mono-jet }
\newcommand{\monoWZ}{mono-$W/Z$(had) }
\newcommand{\monoX}{mono-$X$ }
\newcommand{\monoXnospace}{mono-$X$}
\newcommand{\gq}{g_q}
\newcommand{\gX}{g_{\chi}}
\newcommand{\sqrtgqgX}{\sqrt{g_q g_{\chi}}}
\newcommand{\gqX}{g_{q \chi}}

\newcommand{\sigv}{\langle\sigma v\rangle} 

\newcolumntype{C}[1]{>{\centering\let\newline\\\arraybackslash\hspace{0pt}}m{#1}}       

\title{Collide and Conquer: Constraints on Simplified Dark Matter Models using Mono-$\boldsymbol{X}$ Collider Searches}

\author[a,1]{A. J. Brennan,\note{Corresponding author.}}
\author[a]{M. F. McDonald,}
\author[b]{J. Gramling}
\author[c]{and T. D. Jacques}

\affiliation[a]{ARC Centre of Excellence for Particle Physics at the Terascale \\ School of Physics, The University of Melbourne, Victoria 3010, Australia}
\affiliation[b]{Universit\'{e} de Gen\`{e}ve, Quai E. Ansermet 24, 1211 Gen\`{e}ve 4, Switzerland}
\affiliation[c]{SISSA/ISAS, via Bonomea 265, 34136 Trieste, Italy}

\emailAdd{amelia.brennan@coepp.org.au}
\emailAdd{millie.mcdonald@coepp.org.au}
\emailAdd{johanna.gramling@cern.ch}
\emailAdd{thomas.jacques@sissa.it}

\abstract{The use of simplified models as a tool for interpreting dark matter collider searches has become increasingly prevalent, and while early Run II results are beginning to appear, we look to see what further information can be extracted from the Run I dataset. We consider three `standard' simplified models that couple quarks to fermionic singlet dark matter: an $s$-channel vector mediator with vector or axial-vector couplings, and a $t$-channel scalar mediator. Upper limits on the couplings are calculated and compared across three alternate channels, namely mono-jet, mono-$Z$ (leptonic) and mono-$W/Z$ (hadronic). The strongest limits are observed in the mono-jet channel, however the computational simplicity and absence of significant $t$-channel model width effects in the mono-boson channels make these a straightforward and competitive alternative. We also include a comparison with relic density and direct detection constraints.}

\begin{document}

\hfill
SISSA 18/2016/FISI

\maketitle
\flushbottom

\section{Introduction}
\label{sec:sec1}
\input{Introduction.tex}

\section{Simplified model phenomenology}
\label{sec:sec2}
\input{Models.tex}

\section{Recasting mono-$\boldsymbol{X}$ constraints}
\label{sec:sec3}
\input{Reanalysis.tex}

\section{Results and discussion}
\label{sec:sec4}
\input{Results.tex}

\section{Conclusion}
\label{sec:sec5}
\input{Conclusion.tex}

\section{Acknowledgements}
\label{sec:sec6}
\input{Acknowledgements.tex}

\appendix

\section{Limit setting strategy}
\label{Appendix_limitsetting}
\input{AppendixA.tex}

\section{Validation of signal simulation and event selection procedures}
\label{Appendix_validation}
\input{AppendixB.tex}

\end{document}

%% file: Introduction.tex
Simplified models have emerged as a powerful tool for the interpretation of collider, direct and indirect detection signals of dark matter (DM). Previously, ATLAS and CMS searches for DM were conducted within the context of both Effective Field Theories (EFTs) \cite{Aad:1363019, ATLAS-CONF-2012-147, CMS-PAS-EXO-12-048, Abdallah:1472683} and full UV-complete theories such as Supersymmetry \cite{Aad:2012ms, Aad:2012fqa, Aad:2014wea, SUSY_official_paper}. The latter approach, though well-motivated, is typified by a broad parameter space and generally yields results which are insensitive to the wider class of DM models. EFT constraints, in comparison, are applicable to a broad range of models and rely on the specification of only a small set of parameters, namely the suppression scale, $\Mstar$, and the DM mass, $\mDM$ \cite{DMCons2}.

In the EFT framework, interactions between the dark and Standard Model (SM) sectors are parametrised by a set of higher-dimensional effective operators that arise when the mass of the mediating particle is assumed to be significantly larger than the momentum transferred in a given interaction. Where this condition is not fulfilled, the EFT prescription can produce constraints which detour dramatically from those of the associated UV-complete model \cite{Bai:2010hh, DMCons2, Fox:2011fx, Graesser:2011vj, An:2011ck}. This is less important in direct detection experiments where the momentum transferred in the scattering of DM particles with heavy nuclei is generally of the order of tens of MeV \cite{EFTDM, DMCons3}, or in indirect searches where the annihilations of non-relativistic DM particles in the galactic halo occur with momentum transfers of order $\mDM$. However, for hadron collider searches, where the accessible center of mass energy of two colliding baryons may be sufficient to produce the mediator on-shell, the range of validity of the EFT prescription is significantly diminished. Indeed, recent works have quantitatively shown that the EFT approach can break down as a valid interpretation of data collected during the $\sqrt{\hat{s}} =$ 8 TeV Run I of the Large Hadron Collider (LHC) \cite{Buchmueller:2013dya,ValidEFT, ValidEFT_part2, ValidEFT_part3}. In light of this, simplified models have been investigated as an alternative approach.

In a nutshell, a simplified model (SiM) arises when the heavy mediator which was integrated out in the EFT framework is reintroduced. This must be done with caution in order to ensure that the phenomenology of the resultant SiM is representative of a realistic UV-complete theory of DM.
Like EFTs, SiMs facilitate the comparison of results obtained in  different avenues of dark matter study \cite{DiFranzo:2013vra, Buckley:2014fba} and are defined by a relatively small set of parameters - often $\mDM$, the mass of the mediator $\Mmed$, and the SM-mediator and DM-mediator coupling strengths, $\gq$ and $\gX$ (or $\gqX$ in the case of a single, SM-DM-mediator coupling). This increased parameter space makes it challenging to scan as wide a range of models as can be done with EFTs. However, constraints calculated within the context of a SiM are valid across a comparatively broader energy range.
Whilst EFTs remain useful if used with care, simplified models have become the preferred tool for the interpretation of collider DM searches \cite{DM_MET_LHC, DMOxfordReport, DMForumReport, Harris:2014hga,Buchmueller:2014yoa}.

In this paper, we examine a phenomenologically distinct set of SiMs. In particular, we place constraints on the SiMs corresponding to the simplest UV-completions of the D5 (vector) and D8 (axial-vector) effective operators in the $s$-channel\footnote{The D5 and D8 operators form a nice starting point in the analysis of SiMs as they have been studied exhaustively in the past (see refs.~\cite{Aad:1363019, ATLAS-CONF-2012-147, CMS-PAS-EXO-12-048, Buckley:2013jwa, Aad:2015zva, MonoX, ValidEFT, ValidEFT_part2, ValidEFT_part3} among others). This attention is motivated by the fact that collider limits for the D5 (D8) operator can be readily transformed into limits on spin-independent (spin-dependent) DM-nucleon scattering and vice versa.}. We also include a case in which a scalar mediator is exchanged in the $t$-channel, motivated by its analogue of squark exchange in Supersymmetry. In the heavy mediator limit, the operator describing this model can be rearranged via a Fierz transformation into a combination of operators D5 to D8.

The models are constrained using public results from \monoX + missing transverse energy ($\met$) searches conducted by the ATLAS Collaboration. Specifically, we focus on searches where $X$ is either a parton (manifesting in the detector as a narrow-radius jet), a leptonically-decaying $Z$ boson (manifesting as two opposite-sign same-flavor leptons), or a hadronically-decaying $W$ or $Z$ boson (manifesting as a large-radius jet). The purpose of this work is to strengthen existing SiM limits using the full 20.3 $fb^{-1}$ of Run I ATLAS data, and to explore an enhanced phase space with respect to the mediator and DM masses and the relative strength of the couplings to the visible and dark sectors. We choose to treat the mediator width as the minimal value naturally arising, which is more realistic than a fixed width. Lastly, we provide a cross-check and comparison of the performance of the three targeted collider detection channels, and compare against relic density and direct detection constraints.

The remainder of the paper is organised as follows. Section \ref{sec:sec2} contains a compendium of the SiMs chosen for analysis and the associated collider phenomenology. Section \ref{sec:sec3} outlines the techniques used to recast \monoX + $\met$ limits on the visible cross-section for any new physics process into constraints on SiMs, and specifically on the couplings $\gq$ and $\gX$. Lastly, our results are presented in section \ref{sec:sec4} along with a discussion of the implications of this work. Appendices \ref{Appendix_limitsetting} and \ref{Appendix_validation} include  details of the limit setting and analysis validation procedures.

%% file: Models.tex
\subsection{Model descriptions}
We begin with a short set of assumptions: that the DM particle, $\chi$, is a weakly interacting Dirac fermion, that it is a singlet under the SM, and that it is the lightest stable new particle.
Additionally, the new sector is assumed to couple only to the SM quarks. While possible couplings to SM leptons or gluons have been studied elsewhere (see, for example, ref.~\cite{Fox:2011fx, SiM_gluons}), they are beyond the scope of this paper. The nature of the mediating particle then results from these assumptions: in the $s$-channel it is chosen to be a vector particle which then must also be a SM singlet, denoted $\xi$, while in the $t$-channel we choose a scalar particle which is necessarily charged and coloured, and labelled $\phi$.

The $s$-channel models chosen for analysis are $Z'$-type models characterised by vector ($sV$) or axial-vector ($sA$) couplings to both the dark and SM sectors. They are described by the following interaction Lagrangians:
\begin{equation}
\label{L_int_sV}
\mathcal{L}_{sV} \supset - \xi_{\mu}\left[ \sum\limits_{q} \gq\bar{q}\gamma^{\mu}q + \gX\bar{\chi}\gamma^{\mu}\chi\right],
\end{equation}
\begin{equation}
\label{L_int_sA}
\mathcal{L}_{sA} \supset - \xi_{\mu}\left[\sum\limits_{q} \gq\bar{q}\gamma^{\mu}\gamma_{5}q + \gX\bar{\chi}\gamma^{\mu}\gamma_{5}\chi\right],
\end{equation}
where the sum is over all quarks. This is a simple extension of the standard model and has been studied extensively \cite{Buchmueller:2014yoa, Heisig:2015ira,Blennow:2015gta,Lebedev:2014bba, Alves:2015pea, Alves:2013tqa, Alves:2015mua, An:2012va, An:2012ue, Frandsen:2012rk, Arcadi:2013qia, Shoemaker:2011vi, Frandsen:2011cg, Gondolo:2011eq, Fairbairn:2014aqa, Harris:2014hga, NordstromSVD, Bell:2015rdw, Chala:2015ama, Kahlhoefer:2015bea}.
For the couplings $\gq$ and $\gX$ to remain within the perturbative regime, they are required to satisfy $\gq,\gX \leq 4\pi$. Stronger criteria do exist, though perturbativity breaks down due to the width becoming large before this becomes a problem. Our treatment of the width is discussed in sec.~\ref{width_effects}.

The $t$-channel model (abbreviated $tS$) is primarily motivated by analogy with a common aspect of Supersymmetric models: neutralino DM interacting with the SM sector via $t$-channel exchange of a squark \cite{SUSYDM}, and has been studied within the context of collider searches by a number of groups \cite{DiFranzo:2013vra, Bai:2013iqa, An:2013xka, Chang:2013oia, Zurek:tchannel, Garny:2015wea,  Garny:2014waa, Bell:2011if, Bell:2015rdw}. Note that in this Supersymmetric scenario the DM particle is a Majorana fermion. The collider phenomenology of a SiM in which $\chi$ is of Majorana type is kinematically identical to the corresponding Dirac case (requiring multiplication of the cross-section by a simple factor in order to compute limits) and so Majorana DM is not covered here\footnote{The exception being in the validation of the \monoZ channel, see Sec. \ref{monoZ_validation}.}.

In the $tS$ model, the mediator is allowed to couple to either the left or right-handed quarks as an SU(2) doublet or singlet respectively. Since the LHC is insensitive to the chirality of the quarks, we assume for simplicity that $\phi$ couples to the left-handed quarks only, and is itself an SU(2) doublet, allowing radiation of a $W$ boson. To avoid different couplings to quarks of different generations, and to remain in step with the DM Forum recommendations \cite{DMForumReport}, we include three generations of mediator doublets $\phi_i$, with equal masses and couplings. The interaction Lagrangian for this model is then:
\begin{equation}
\label{L_int_tS}
\mathcal{L}_{tS} \supset \sum_{i} \gqX \bar{Q_i} P_R \phi_i \chi + {\rm h.c.},
\end{equation}
where the sum is over the three quark doublets, $\gqX$ is the DM-quark coupling (equal for each generation), and $P_R$ is the usual chiral projection operator.

\subsection{The mono-$\boldsymbol{X}$ + $\boldsymbol{\met}$ signatures}
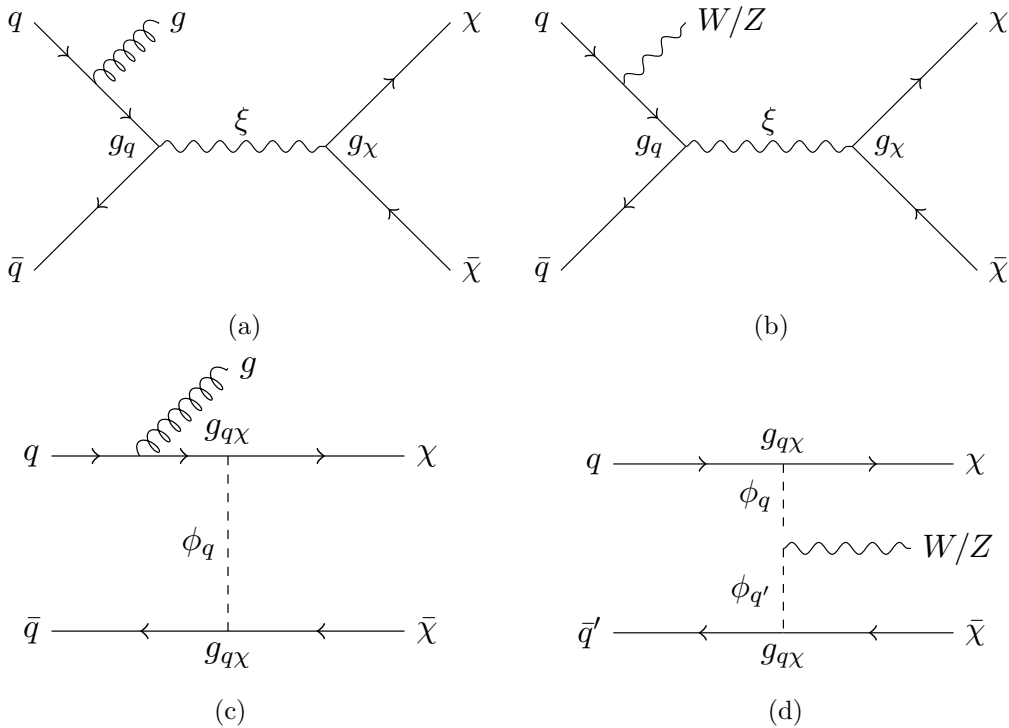
\begin{figure}[t]
  \centering
  \begin{subfigure}[b]{0.45\textwidth}
    \centering
    \resizebox{\linewidth}{!}{
      \begin{tikzpicture}
        \draw[fermion] (-1.5,1.5)node[left]{$q$} --(-0.75,0.75);
        \draw[gluon] (-0.75,0.75) -- (0,1.5)node[right]{$g$};
        \draw[fermion] (-0.75,0.75) -- (0,0);
        \draw[antifermion] (-1.5,-1.5)node[left]{$\bar{q}$} --(0,0);
        \draw[fill] (0,0) circle [radius=0.0]node[left]{$\gq\mbox{ }$};
        \draw[photon] (0,0) --node[above]{$\xi$} (2,0);
        \draw[fermion] (2,0) -- (3.5,1.5)node[right]{$\chi$};
        \draw[antifermion] (2,0) --(3.5,-1.5)node[right]{$\bar{\chi}$};
        \draw[fill] (2,0) circle [radius=0.0]node[right]{$\mbox{ }\gX$};
      \end{tikzpicture}
    }
    \caption{}
    \label{fig:FD_sV_gluonISR}
  \end{subfigure}
  \begin{subfigure}[b]{0.45\textwidth}
    \centering
    \resizebox{\linewidth}{!}{
      \begin{tikzpicture}
        \draw[fermion] (-1.5,1.5)node[left]{$q$} --(-0.75,0.75);
        \draw[photon] (-0.75,0.75) -- (0,1.5)node[right]{$W/Z$}; 
        \draw[fermion] (-0.75,0.75) -- (0,0);
        \draw[antifermion] (-1.5,-1.5)node[left]{$\bar{q}$} --(0,0);
        \draw[fill] (0,0) circle [radius=0.0]node[left]{$\gq\mbox{ }$};
        \draw[photon] (0,0) --node[above]{$\xi$} (2,0);
        \draw[fermion] (2,0) -- (3.5,1.5)node[right]{$\chi$};
        \draw[antifermion] (2,0) --(3.5,-1.5)node[right]{$\bar{\chi}$};
        \draw[fill] (2,0) circle [radius=0.0]node[right]{$\mbox{ }\gX$};
      \end{tikzpicture}
    }
    \caption{}
    \label{fig:FD_sV_WZISR}
  \end{subfigure}
  \begin{subfigure}[b]{0.4\textwidth}
    \centering
    \resizebox{\linewidth}{!}{
      \begin{tikzpicture}
        \draw[fermion] (-2,1.)node[left]{$q$} --(-1,1.);
        \draw[fermion] (-1,1.) --(0,1.);
        \draw[gluon] (-1,1.) --(0,2.)node[right]{$g$};
        \draw[antifermion] (-2,-1)node[left]{$\bar{q}$} --(0,-1);
        \draw[fill] (0,1.) circle [radius=0.0]node[above]{$\gqX$};
        \draw[dashed] (0,1.) -- node[left]{$\phi_{q}$}(0,-1);
        \draw[fermion] (0,1.) -- (2,1.)node[right]{$\chi$};
        \draw[antifermion] (0,-1) --(2,-1)node[right]{$\bar{\chi}$};
        \draw[fill] (0,-1) circle [radius=0.0]node[below]{$\gqX$};
      \end{tikzpicture}
    }
    \caption{}
    \label{fig:FD_tS_gluonISR}
  \end{subfigure}
  \hspace{1cm}
  \begin{subfigure}[b]{0.4\textwidth}
    \centering
    \resizebox{\linewidth}{!}{
      \begin{tikzpicture}
        \draw[fermion] (-2,1.)node[left]{$q$} --(0,1.);
        \draw[antifermion] (-2,-1)node[left]{$\bar{q}'$} --(0,-1);
        \draw[fill] (0,1.) circle [radius=0.0]node[above]{$\gqX$};
        \draw[dashed] (0,1.) -- node[left]{$\phi_{q}$}(0,0.25);
        \draw[photon] (0,0.) -- (1.5, 0.)node[right]{$W/Z$};
        \draw[dashed] (0,0.) -- node[left]{$\phi_{q'}$}(0,-1);
        \draw[fermion] (0,1.) -- (2,1.)node[right]{$\chi$};
        \draw[antifermion] (0,-1) --(2,-1)node[right]{$\bar{\chi}$};
        \draw[fill] (0,-1) circle [radius=0.0]node[below]{$\gqX$};
      \end{tikzpicture}
    }
    \caption{}
    \label{fig:FD_tS_WZmediator}
  \end{subfigure}
  \caption{A representative subset of dark matter pair-production processes with a gluon or $W/Z$ boson in the final state for the $s$-channel (a,b) and $t$-channel (c,d) models. Note that other diagrams are possible, including initial state radiation of a gauge boson, and internal bremsstrahlung of a gluon.}
  \label{allchannel_sig_phen}
\end{figure}

The \monoX + $\met$ signal (abbreviated to \monoXnospace) is a popular collider signal in the search for new physics, particularly in the search for dark matter. Since DM particles are not expected to interact with detector material, they appear as missing transverse energy when balanced against a visible object, $X$, that is radiated from the initial or intermediate state. For the $s$-channel SiMs discussed above, only initial-state radiation is permitted; see figs.~\ref{fig:FD_sV_gluonISR} and \ref{fig:FD_sV_WZISR} for examples. For the $tS$ model, radiation of a gluon or electroweak (EW) boson is permitted both from initial state partons (fig.~\ref{fig:FD_tS_gluonISR}) or from the mediator (fig.~\ref{fig:FD_tS_WZmediator}). Note that these diagrams do not comprise a comprehensive set.

The most likely scenario at the LHC is the production of a jet alongside the invisible $\chi$ pair, as a result of the strong coupling and prevalence of partons in the initial state. However, to fully exploit the potential of the ATLAS detector to record and identify a vast array of particle types, we also consider two additional channels. Firstly, we take advantage of the relative cleanliness and simplicity of leptons in the leptonically-decaying mono-$Z$ ($\rightarrow \ell^+ \ell^-)$ channel. We also take advantage of the large hadronic branching fraction, and developing jet-identification techniques for boosted EW bosons, in the hadronically-decaying mono-$W/Z$ ($\rightarrow jj)$ channel\footnote{In addition, one of the first Run II dark matter search results from ATLAS was from this channel \cite{monoWZ_run2}, released during the preparation of this paper.}. In both cases, the large multi-jet background is reduced, and complications in jet production such as parton-matching can be ignored, making these an interesting alternative to the \monojet channel where speed, efficiency and a reduction in jet-associated uncertainties may make up for a loss in sensitivity.

\subsection{Mass and coupling points}
A representative set of dark matter and mediator masses, listed in table \ref{Mass_coup_points}, are chosen for study in each detection channel. DM masses of 3, 30 and 300 GeV are also included in the \monoZ channel, where ease of production permits higher granularity in the mass phase space. All $(\mX, \Mmed)$ combinations are allowed in the $sV$ and $sA$ models, while in the $tS$ model $\Mmed$ must be greater than $\mX$ to ensure stability of the DM particle. The couplings $\gq$ and $\gqX$ are set to unity, while the DM-mediator coupling in the $s$-channel models, $\gX$, is varied from 0.2 to 5. The mediator masses are chosen to cover a broad range of parameter space and to coincide with predominantly three regimes: (near-)degenerate ($\Mmed \approx \mX$), on-shell ($\Mmed \geq 2 \mX$) and off-shell ($\Mmed < 2 \mX$).

\begin{table}
\centering
\begin{tabular}{C{3cm} | C{3cm} | C{1.5cm}  C{1.5cm} | C{3cm}}
\hline
\hline
\multirow{2}{*}{$\mX$ [GeV]} & \multirow{2}{*}{$\Mmed$ [GeV]} & \multicolumn{2}{c|} {$s$-channel} & $t$-channel \T \B \\
& & $\gq$ & $\gX$ & $\gqX$ \T \B\\
\hline
1, (3), 10, (30), 100, (300), 1000 & 1, 2, 10, 20,  100, 200, 1000, 2000 & 1 & 0.2, 0.5, 1, 2, 5 & 1 \T \B  \\
\hline
\hline
\end{tabular}
\caption{Mass and coupling points chosen for the analysis of simplified dark matter models. Values in brackets are only included in the \monoZ channel. The mediator masses are primarily representative of three regimes: (near-)degenerate ($\Mmed \approx \mX$), on-shell ($\Mmed \geq 2 \mX$) and off-shell ($\Mmed < 2 \mX$). For the $t$-channel model, $\Mmed > \mX$ is required to ensure stability of the DM particle.}
\label{Mass_coup_points}
\end{table}

\subsection{Treatment of the width}
\label{width_effects}
An important factor when considering SiMs is to ensure that the mediator width is treated appropriately, as it impacts both the cross-section calculation and, in some cases, the kinematic behaviour of the model.

Following the DM Forum recommendations \cite{DMForumReport}, we use the minimal width, allowing coupling to all kinematically accessible quarks. We assume minimal flavour violation, which implies a universal coupling to all quark flavours. The minimum width for each model is given by\footnote{It is possible that the mediator may decay to other SM or BSM particles \cite{Harris:2014hga}, but this is not expected to have a large effect on the kinematic distribution as long as the width remains relatively small \cite{DMForumReport}.}:

\begin{eqnarray}
    \Gamma_{sV} \, &=& \,  \frac{\gX^2 M}{12\pi}\left(1 + \frac{2 \mX^{2}}{M^{2}}\right)\left(1 - \frac{4 \mX^{2}}{M^{2}}\right)^{\frac{1}{2}} \Theta(M-2 \mX) \nonumber\\
                  && + \sum_{\substack{q}}\frac{\gq^2M}{4\pi}\left(1 + \frac{2m_{q}^{2}}{M^{2}}\right)\left(1 - \frac{4m_{q}^{2}}{M^{2}}\right)^{\frac{1}{2}} \Theta(M-2m_q)\\[5pt]
    \Gamma_{sA} \, &=& \,  \frac{\gX^2 M}{12\pi}\left(1 - \frac{4 \mX^{2}}{M^{2}}\right)^{\frac{3}{2}} \Theta(M-2 \mX) \nonumber\\
                  && + \sum_{\substack{q}}\frac{\gq^2 M}{4\pi}\left(1 - \frac{4m_{q}^{2}}{M^{2}}\right)^{\frac{3}{2}} \Theta(M-2m_q) \\[5pt]
    \Gamma_{tS} \, &=& \,  \sum_{\substack{q}} \frac{\gqX^2M}{16\pi}\left(1 - \frac{m_{q}^{2}}{M^{2}} - \frac{\mX^{2}}{M^{2}}\right) \nonumber\\
                  && \times \sqrt{\left(1 - \frac{m_{q}^{2}}{M^{2}} + \frac{\mX^{2}}{M^{2}}\right)^{2} - 4\frac{\mX^{2}}{M^{2}}} \,\, \Theta(M-m_q- \mX)
\end{eqnarray}

\begin{figure}[t]
  \begin{center}
    \includegraphics[width=0.495\textwidth]{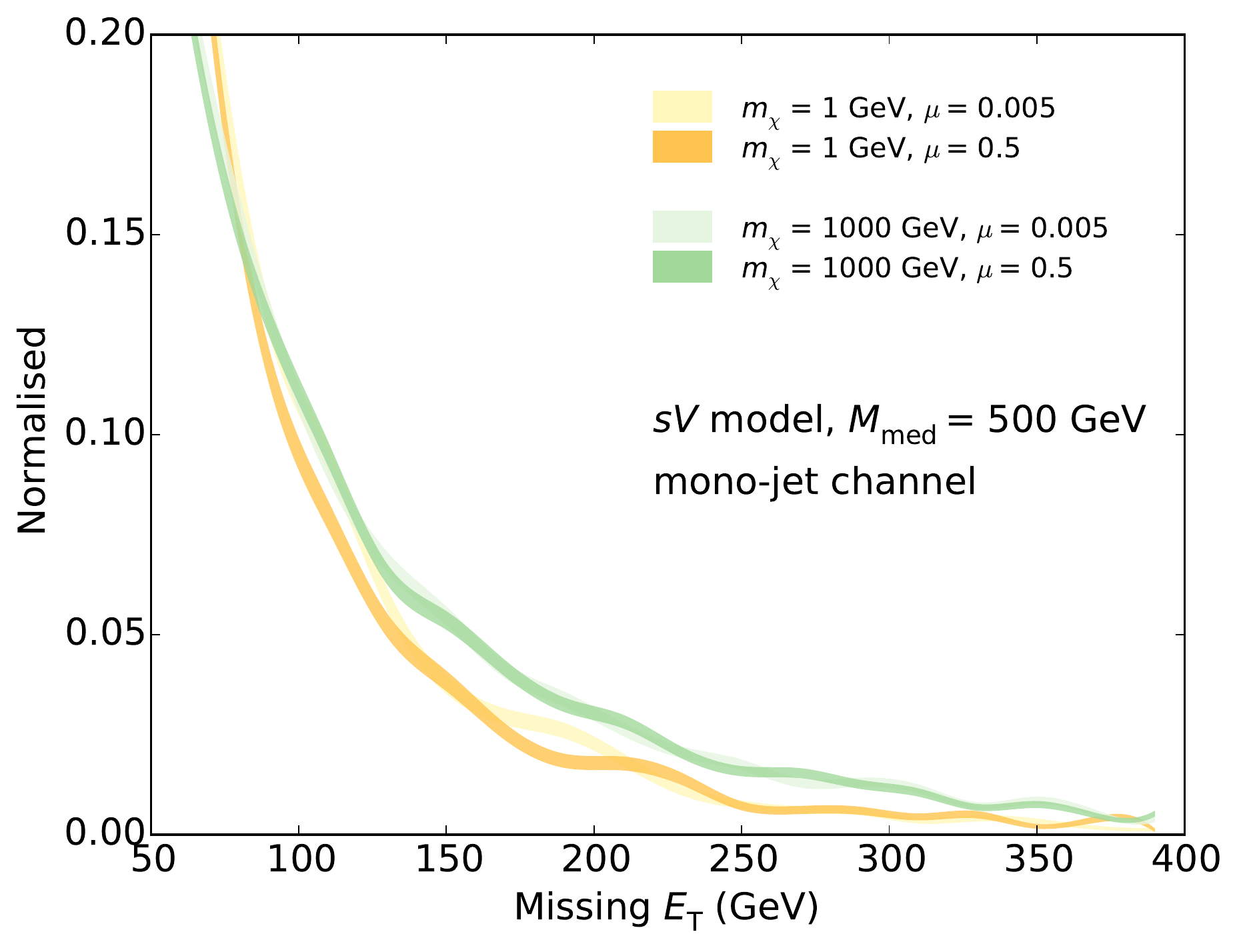}
    \includegraphics[width=0.495\textwidth]{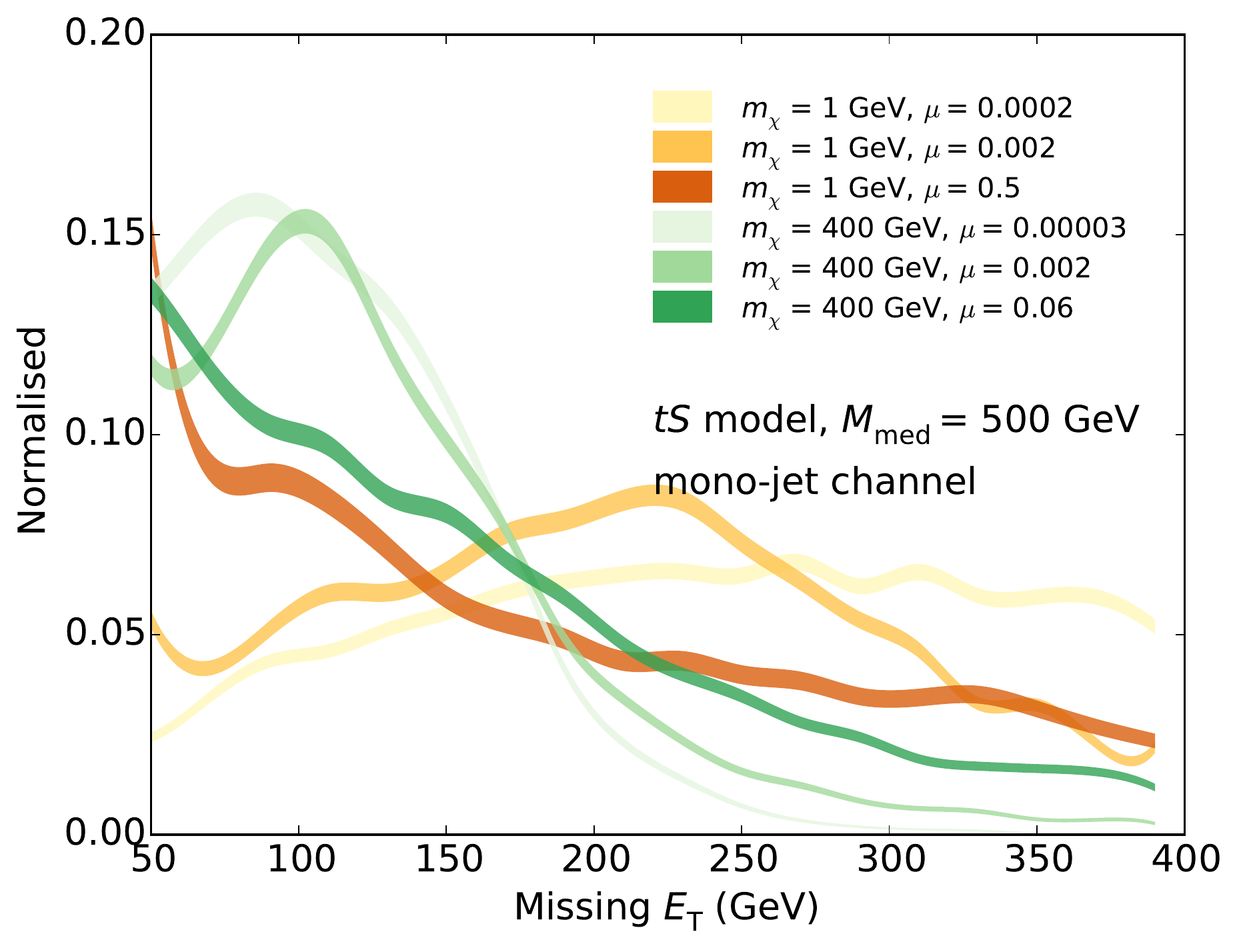}
    \includegraphics[width=0.495\textwidth]{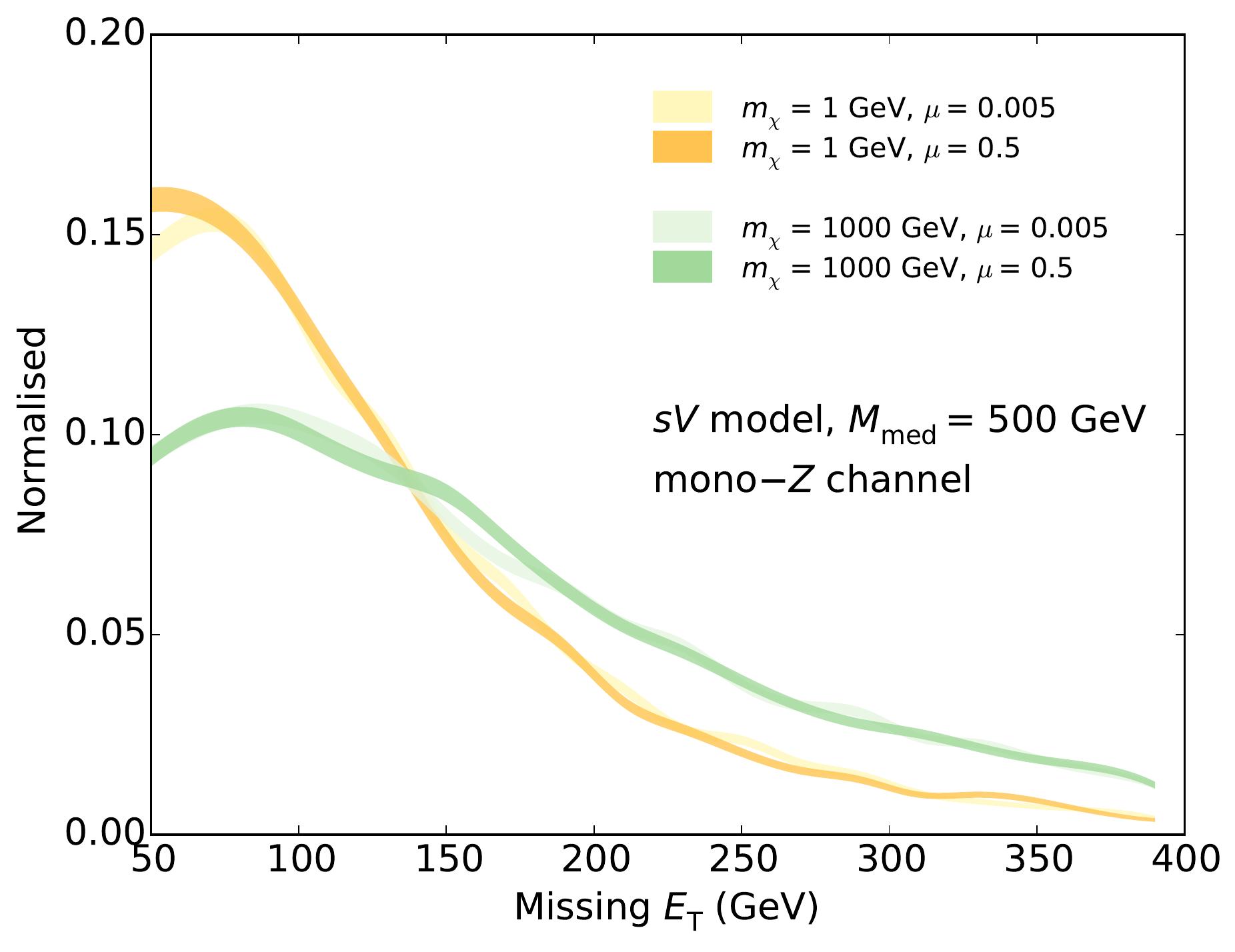}
    \includegraphics[width=0.495\textwidth]{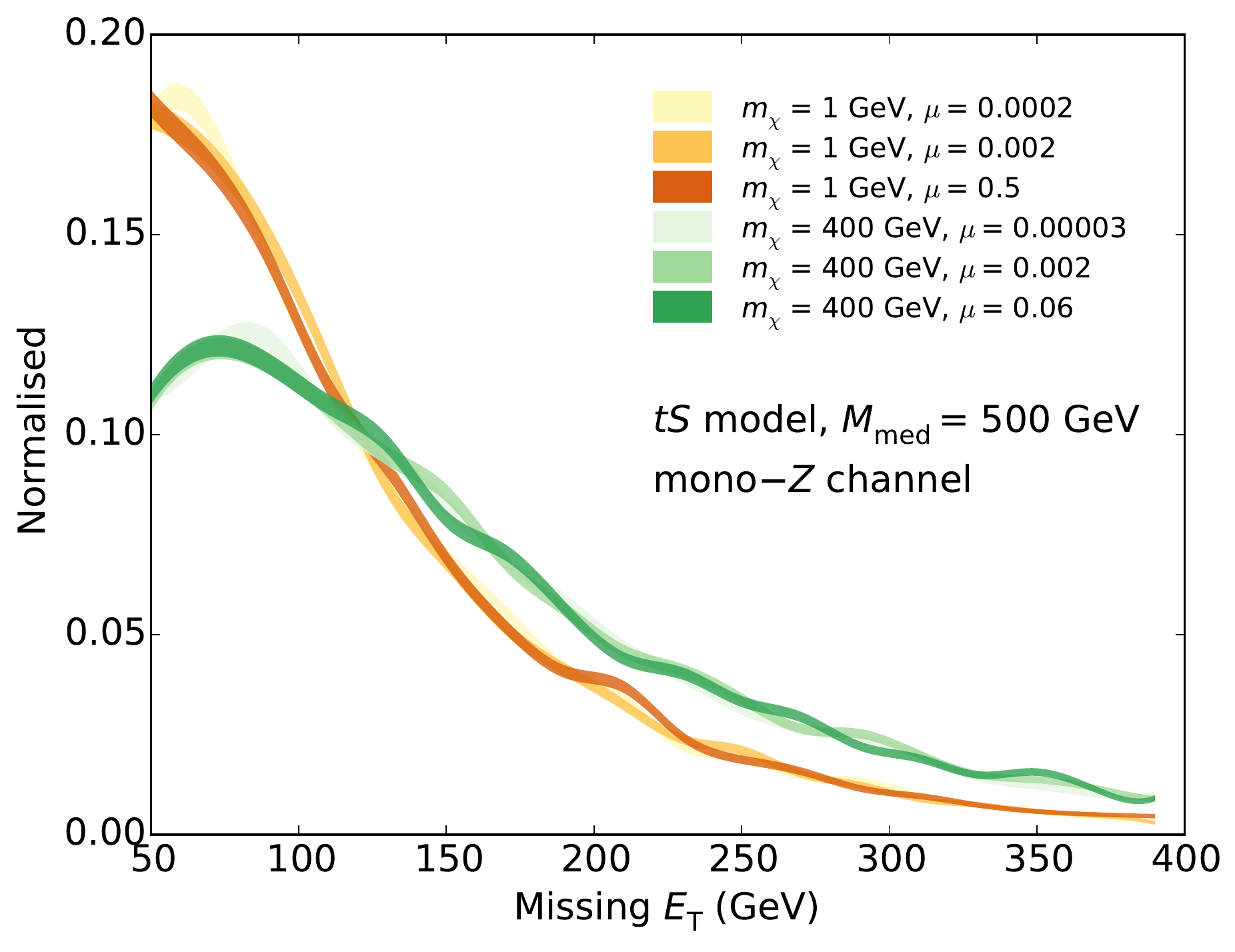}
    \caption{The $\met$ distribution of the $sV$ and $tS$ models in the \monojet and \monoZ channels, for some example masses. The parameter $\mu$ is defined as $\Gamma / \Mmed$, and is used to demonstrate the impact of a changing width; the $tS$ model in the \monojet channel shows a clear width dependence, while all other model/channel combinations show behaviour that is independent of the width for the phase space considered. The widths are obtained with couplings of 0.1, 1 and 5 where $\mu < 0.5$ remains true.}
    \label{fig:MET_dists}
  \end{center}
\end{figure}

We can take advantage of the fact that for each point in ($\mX$, $\Mmed$) phase space, the mediator width (and therefore the couplings) do not greatly affect a model's kinematic behaviour (with the notable exception of the $tS$ model in the \monojet channel). This is demonstrated in fig.~\ref{fig:MET_dists}, where we plot a simplified $\met$ distribution (as a proxy for the full selection in each analysis) for the $sV$ (representing both the $sV$ and $sA$ models) and $tS$ models for two mass points and a demonstrative set of couplings such that $\Gamma < \Mmed/2$. The $\met$ distribution is predominantly independent of the mediator width for the $s$-channel models in the \monojet channel, and all models in the mono-$Z$(lep)\footnote{In this discussion, the \monoWZ channel can be assumed to follow the same logic as for the \monoZ channel.} channel. For the $s$-channel models, the same result was found by ref.~\cite{DMForumReport}, which provides a set of recommendations for the usage of simplified models for DM searches in Run II. As described below, this independence of the kinematic spectrum on the width, and therefore the couplings, allows a simplification of the limit calculation used in sec.~\ref{Appendix_limitsetting}. 

However, there is a clear variation in the kinematic behaviour of the tS model in the \monojet channel, which can be attributed to additional diagrams (accessible only in this channel) featuring a gluon in the initial state and subsequently allowing the mediator to go on-shell. These diagrams are discussed and shown in, for example, ref.~\cite{Zurek:tchannel}. The kinematics of these diagrams with an on-shell mediator are much more sensitive to variations in the width.

In the cases where the kinematic distribution is independent of the width, we assume that the impact of the selection cuts in each channel is unchanged by the couplings. In this case, the following relations approximately hold:

\begin{equation}
  \sigma \propto
  \begin{cases}
      \gq^2 \gX^2 / \Gamma & \mathrm{ if } \, \Mmed \geq 2 \mDM\\
      \gq^2 \gX^2 & \mathrm{ if } \, \Mmed < 2 \mDM
  \end{cases}
  \label{eq:sigma_propto_couplings_schan}
\end{equation}
in the $sV$ and $sA$ models  \cite{NordstromSVD}, and

\begin{equation}
  \sigma \propto \gqX^4
  \label{eq:sigma_propto_couplings_tchan}
\end{equation}
in the $tS$ model. When valid, these approximations allow us to greatly simplify our limit calculations, and for this reason, we restrict our primary results to regions of parameter space where $\Gamma/\Mmed < 0.5$ (see app.~\ref{Appendix_limitsetting} for further details of the limit-setting calculation).

The generator treatment of the mediator as a Breit-Wigner propagator, rather than a true kinetic propagator, breaks down for large widths \cite{An:2012va,NordstromSVD}. More problematically, it was noted by refs.~\cite{NordstromSVD,An:2012va} that the Breit-Wigner propagator breaks down in the $\mDM \gg \Mmed$ region even if $\Gamma/\Mmed$ is small. To correct for this we follow ref.~\cite{NordstromSVD}, and rescale the cross-section in the $\mDM > \Mmed$ region by a factor which takes into account the error introduced by the use of a Breit-Wigner propagator by the generator. The factor is found by convolving the PDF with both the kinetic and Breit-Wigner propagators in turn and taking the ratio at each mass point. We approximate the kinetic propagator by making the substitution $\Mmed \Gamma(\Mmed) \rightarrow s \Gamma(\sqrt{s}) / \Mmed$ in the Breit-Wigner propagator.

A full study of the $tS$ model within the \monojet channel, where altering the coupling can lead to changed kinematic behaviour, has been performed elsewhere \cite{Zurek:tchannel}, and requires the production of individual samples for each coupling point. This, combined with the challenges associated with including differing orders of $\alpha_s$, make the generation process computationally expensive compared to the \monoZ and \monoWZ channels. We therefore exclude an analysis of the $tS$ model in the \monojet channel in this work.

%% file: Reanalysis.tex
The procedure for recasting existing \monoX analyses to obtain SiM constraints follows a simple cut-and-count methodology. Firstly, signal events are simulated (described below in section \ref{signal_generation}) with object $p_{\mathrm{T}}$ smearing applied to approximate the detection efficiency of the ATLAS detector, $\epsilon$. The event selection criteria of the \monoX analysis of interest is then applied to the simulated signal samples. Events surviving the selection criteria are counted to determine the likelihood of a dark matter event being observed (referred to as the acceptance, $\mathcal{A}$), which is then used in combination with channel-specific model-independent limits on new physics events to limit the parameter space of a given model.
For a comprehensive description of the recasting procedure, see appendix \ref{Appendix_limitsetting}.

In this paper, \monojet constraints are derived from a search for new phenomena conducted by the ATLAS Collaboration using $pp$ collisions at $\sqrt{s}=$ 8 TeV as described in ref. \cite{Aad:2015zva}. Similarly, the leptonic mono-$Z$ and hadronic mono-$W/Z$ constraints are derived from ATLAS dark matter searches that were optimised for the D1, D5 and D9 effective operators \cite{Aad:2014monoZlep,Aad:2013monoWZ}. These analyses are described in further detail in sections \ref{monojet_constraints}, \ref{monoZ_constraints} and \ref{monoWZ_constraints} respectively.

\subsection{Signal simulation}
\label{signal_generation}
Monte Carlo simulated event samples are used to model the expected signal for each channel and for each SiM. Leading order matrix elements for the process $pp \rightarrow X + \chi\bar{\chi}$ (where $X$ is specifically one or two jets\footnote{Jets are seeded by any parton excluding the (anti-)top quark.}, a $Z(\rightarrow \ell^+ \ell^-)$ boson or a $W/Z(\rightarrow$ $jj$) boson) are first simulated using \MGnospace$\_$aMC$@$NLO v2.2.2 \cite{MG_aMCNLO2014} with the MSTW2008lo68cl PDF \cite{MSTW}. During this stage, the renormalisation and factorisation scales are set to the default sum of $\sqrt{m^{2} + p_{T}^{2}}$ for all particles in the final state. Showering and hadronisation are then performed by \PYTHIAnospace .201 \cite{pythia8} with the appropriate PDF and using the ATLAS UE Tune AU2-MSTW2008LO~\cite{AUtune}. Reconstruction of small-radius jets (hereon referred to just as `jets') for the \monojet channel is performed by \FASTJET~\cite{FastJet} using the anti-$k_{\mathrm{T}}$ algorithm with radius parameter $R$ = 0.4. Similarly, reconstruction of large-radius jets for the \monoWZ channel is performed using the Cambridge-Aachen algorithm with $R$ = 1.2. The latter channel also includes a mass-drop filtering procedure with $\mu$ = 0.67 and $\sqrt{y}$\footnote{$\sqrt{y} = \mathrm{min}(p_{\mathrm{T}_{j1}},p_{\mathrm{T}_{j2}})\Delta R / m_{jet}$ is the momentum balance of the two leading subjets.} = 0.4 (see ref.~\cite{massdrop} for further details), which favours large-$R$ jets with two balanced subjets, consistent with the decay of an EW boson to a (potentially boosted) dijet pair. Lastly, the detector response is approximated by applying a Gaussian smearing factor to the $p_{\mathrm{T}}$ of all leptons and jets.

\subsubsection{Parton matching scheme}
\label{matching_procedure}
In the ATLAS \monojet analysis, matching of partons generated in \MG to jets generated in \PYTHIA is performed using the MLM scheme \cite{MLMscheme}, with two matching scales, or values of `QCUT', per mass/coupling point. In combination, the QCUT values span a broad kinematic range with a cut placed on the leading jet $p_{\mathrm{T}}$ per event to avoid double-counting. This treatment aims to both enhance the statistics in the high $\met$ signal regions and to mitigate the impact of the matching scale on the shape of the $p_{\mathrm{T}}$ and $\met$ distributions; that is, to reduce the uncertainty in those areas of phase space where the transferred momentum is significantly larger or smaller that the QCUT value. For the analysis of SiMs in this work, we use instead a single matching scale of 80 GeV. The need for a second, high $\met$ QCUT is compensated for by the generation of increased events per mass and coupling sample. Furthermore, any effects introduced by this simplified procedure are accounted for by a conservative estimation of the uncertainties on the final limits as discussed in sec.~\ref{uncertainty_estimation_proc}. Though not ideal, this approach suitably reproduces the results of the ATLAS \monojet analysis for the masses of interest (see sec.~\ref{monojet_validation}). Importantly, it also reduces the complexity and computational expense involved in estimating limits for the \monojet channel.

We now move to a discussion of each of the \monoX channels separately.

\subsection{Mono-jet constraints}
\label{monojet_constraints}
The ATLAS \monojet + $\met$ analysis \cite{Aad:2015zva} was originally designed to set limits on three new physics scenarios, the most relevant of which is the production of WIMP DM within the context of a set of effective operators. The analysis also includes a brief study of a $Z'$ DM model which is analogous to our $sV$ model.

Signal selection is carried out based on at least one hard jet recoiling against missing energy. To ensure that the correct back-to-back jet + $\met$ topology is selected events are required to have a leading jet, $j_{1}$, with $p_{T} >$ 120 GeV and $|\eta| <$ 2.0 satisfying $p_{T}^{j_{1}}/\met >$ 0.5. Surviving events must then fulfill $|\Delta\phi(j,\metvec)|>1.0$, where $j$ is any jet with $p_{T} >$ 30 GeV and $|\eta| <$ 4.5. This criterion reduces the multijet background contribution where the large $\met$ originates mainly from jet energy mismeasurements. Note that there is no upper limit placed on the number of jets per event. The contribution from the dominant background processes, $W/Z+$jets, is managed with a veto on events containing muons or electrons with $p_{T}>$ 7 GeV. Lastly, nine separate signal regions are defined with increasing lower thresholds on $\met$, which range from 150 GeV to 700 GeV as shown in table \ref{monojet_SRs}.

The ATLAS \monojet analysis revealed no significant deviation of observed events from the expected SM backgrounds in the 8 TeV dataset of Run I. Subsequently, model-independent limits on new physics signatures were provided in terms of the visible cross-section, $\sigma\times\mathcal{A}\times\epsilon$; these are listed in table \ref{monojet_SRs}.

\begin{table}[!htbp]
\centering
\begin{tabular}{c|c|c}
 \hline
 \hline
 Signal Region & $\met$ threshold [GeV] & $\sigma \times \mathcal{A} \times \epsilon$ [fb] \\
 \hline
 SR1 & 150 & 726 (935) \\
 SR2 & 200 & 194 (271) \\
 SR3 & 250 & 90 (106) \\
 SR4 & 300 & 45 (51) \\
 SR5 & 350 & 21 (29) \\
 SR6 & 400 & 12 (17) \\
 SR7 & 500 & 7.2 (7.2) \\
 SR8 & 600 & 3.8 (3.2) \\
 SR9 & 700 & 3.4 (1.8) \\
 \hline
 \hline
\end{tabular}
\caption{The ATLAS \monojet $\met$ signal regions and corresponding observed (expected) model-independent upper limits on $\sigma \times \mathcal{A} \times \epsilon$ at 95\% confidence level. Adapted from ref. \cite{Aad:2015zva}.}
\label{monojet_SRs}
\end{table}

The signal simulation procedure outlined in sec. \ref{signal_generation} and implementation of the selection criteria discussed above were validated for the \monojet channel via reproduction of ATLAS limits on the suppression scale, $\Mstar \equiv \Mmed / \sqrtgqgX$, for the $Z'$ model. The details of this process are contained in appendix \ref{monojet_validation}. Importantly, we observe agreement within $\sim$12\% for all samples.

\subsection{Mono-$Z$(lep) constraints}
\label{monoZ_constraints}
The ATLAS \monoZ + $\met$ analysis \cite{Aad:2014monoZlep} was principally designed to constrain a set of EFT models of DM. As a secondary focus, it also included a short study of a $t$-channel SiM similar to our $tS$ model.

The selection criteria for this analysis are summarised as follows (see the paper for a full description). Electrons (muons) are required to have a $p_{\mathrm{T}}$ greater than 20 GeV, and $|\eta|$ less than 2.47 (2.5). Two opposite-sign, same-flavour leptons are selected, and required to have invariant mass and pseudorapidity such that $m_{\ell \ell} \in [76, 106]$ GeV and $|\eta^{\ell \ell}| < 2.5$. The reconstructed $Z$ boson should be approximately back-to-back and balanced against the $\met$, ensured with the selections $\Delta \phi (\metvec, p_{\mathrm{T}}^{\ell \ell}) > 2.5$ and $| p_{\mathrm{T}}^{\ell \ell} - \met | \, /  \, p_{\mathrm{T}}^{\ell \ell} < 0.5$. Events containing a jet with $p_{\mathrm{T}}>$ 25 GeV and $|\eta|< $ 2.5 are vetoed. Events are also vetoed if they contain a third lepton with $p_{\mathrm{T}}>$ 7 GeV. The signal regions are defined by increasing lower $\met$ thresholds: $\met >$ 150, 250, 350, 450 GeV.

A cut-and-count strategy is used to estimate the total observed yields and expected SM backgrounds in each signal region. The limits on $\sigma\times\mathcal{A}\times\epsilon$ are not publicly available, so we take the numbers of expected and observed events from ref.~\cite{Aad:2014monoZlep}, along with the associated uncertainties, and convert these into model-dependent upper limits with a single implementation of the HistFitter framework \cite{HistFitter} using a frequentist calculator and a one-sided profile likelihood test statistic (the LHC default). The results of this process are displayed in table~\ref{tab:sigmalim_monoZ}. Note that we use signal regions 1 and 2 only, as our simplified HistFitter approach is inadequate to handle the very low statistics of signal regions 3 and 4. These upper limits, the \monoZ signal generation and the selection procedures are all validated through comparison of the ATLAS analysis limits on a variant of the tS model with our own limits on the same model; see sec.~\ref{monoZ_validation} for details.

\begin{table}[!htbp]
  \begin{center}
    \begin{tabular}{c|c|c}
      \hline
      \hline
      Signal Region & $\met$ threshold [GeV] & $\sigma \times \mathcal{A} \times \epsilon$ [fb] \\
      \hline
      SR1 & 150 & 1.59 (1.71) \\
      SR2 & 250 & 0.291 (0.335) \\
      \hline
      \hline
    \end{tabular}
  \end{center}
  \caption{The ATLAS \monoZ + $\met$ signal regions and corresponding observed (expected) model-independent upper limits on $\sigma \times \mathcal{A} \times \epsilon$ at 95\% confidence level, where those limits have been calculated in this work with HistFitter from the numbers of expected and observed events published in ref.~\cite{Aad:2014monoZlep}.}
  \label{tab:sigmalim_monoZ}
\end{table}

\subsection{Mono-$W/Z$(had) constraints}
\label{monoWZ_constraints}

The ATLAS \monoWZ + $\met$ search \cite{Aad:2013monoWZ} was aimed at constraining the spin-independent effective operators C1, D1, and D5, and the spin-dependent operator D9. The search was originally designed to exploit what was thought to be the constructive interference of $W$ boson emission from opposite-sign up-type and down-type quarks, leading to DM production wherein the mono-$W$ channel is dominant. Recent studies \cite{Bell:gaugeInv} have revealed this scenario to violate gauge invariance and so we ignore it in this analysis.

The \monoWZ event selection is carried out as follows. Large-radius jets are selected using a mass-drop filtering procedure (see sec.~\ref{signal_generation}) to suppress non-$W/Z$ processes. Events are required to contain at least one large-$R$ jet with $p_{\mathrm{T}} >$ 250 GeV, $|\eta| <$ 1.2 and a mass, $m_{\mathrm{jet}}$, within a 30-40 GeV window of the $W/Z$ mass (i.e. $m_{\mathrm{jet}} \in [50, 120]$ GeV). In order to reduce the $t \bar{t}$ and multijet backgrounds, a veto removes events containing a small-$R$ jet with $\Delta\phi(\mathrm{jet},\met)< 0.4$, or containing more than one small-$R$ jet with $p_{\mathrm{T}} >$ 40 GeV, $|\eta| <$ 4.5, and $\Delta R$(small-$R$ jet, large-$R$ jet)$>0.9$. Electrons, muons and photons are vetoed if their $p_{\mathrm{T}}$ is larger than 10 GeV and they lie within $|\eta| <$ 2.47 (electrons), 2.5 (muons), 2.37 (photons). Two signal regions are defined with $\met > 350$ GeV and $\met > 500$ GeV.

The ATLAS analysis used a shape-fit of the mass distribution of the large-$R$ jet to set exclusion limits, however we use the published numbers of SM background and observed data events (along with the associated uncertainties) \cite{Aad:2013monoWZ} to convert to upper limits on new physics events using the HistFitter framework. For the $\met > 500$ GeV signal region, we obtain the limits shown in table~\ref{tab:sigmalim_monoWZ}; these are validated, along with the signal generation and selection process, in sec.~\ref{monoWZ_validation}. We do not consider the first signal region with $\met > 350$ GeV in the recasting procedure, since the cut-and-count limits extracted could not be convincingly validated. The high $\met$ signal region was found to be optimal for most operators studied by the ATLAS analysis.

\begin{table}[!htbp]
  \begin{center}
    \begin{tabular}{c|c|c}
      \hline
      \hline
      Signal Region & $\met$ threshold [GeV] & $\sigma \times \mathcal{A} \times \epsilon$ [fb] \\
      \hline
      SR2 & 500 & 1.35 (1.34) \\
      \hline
      \hline
    \end{tabular}
  \end{center}
  \caption{The ATLAS \monoWZ $\met$ signal region considered in this work and corresponding observed (expected) model-independent upper limits on $\sigma \times \mathcal{A} \times \epsilon$ at 95\% confidence level, where those limits have been adapted from the numbers of expected and observed events in ref.~\cite{Aad:2013monoWZ} using HistFitter.}
  \label{tab:sigmalim_monoWZ}
\end{table}

%% file: Results.tex
\subsection{Limits on the couplings $\boldsymbol{\sqrt{g_{q}g_{\chi}}}$}
The 95\% confidence level upper limits on the $sV$ and $sA$ model coupling combination $\sqrtgqgX$, and the $tS$ model coupling $\gqX$, obtained from each of the \monoX channels, are presented in figs.~\ref{fig:results_sVsA_rat05}-\ref{fig:results_tS}. These quantities were evaluated as described in appendix \ref{Appendix_limitsetting} (including statistical and systematic uncertainties), and correspond to the best limits of each signal region tested.

In each plot, limits are shown ranging from $<$0.01 to the upper perturbative limit for each coupling, $4\pi$; where a limit was calculated to be larger than this, the limit is considered meaningless and the region is coloured grey. The white (hatched) regions coincide with those mass points which yield an initial (final) value of $\sqrtgqgX$ or $\gqX$ which fails to satisfy $\Gamma < \Mmed / 2$. (We observe that values for which the width is just within our upper validity bound of $\Mmed/2$ may be pushed over into the invalid range with the addition of new particles, not considered here, which would serve to increase the mediator width.) When $\gX / \gq$ = 0.2, only the \monojet channel produces limits which survive this requirement, and so these are shown separately in fig~\ref{fig:results_sVsA_rat02}.

Detailed comments specific to each channel are provided below, however some trends are channel-independent. For the $sV$ model, strong limits exist when $\Mmed > 2 \mX$ as the mediator can go on-shell, thereby enhancing the cross-section. The $sA$ model limits are generally similar to the $sV$ model limits except in the region corresponding to $\mX \gtrsim \sqrt{4\pi} \Mmed / g^{\rm gen}_\chi$ where $g^{\rm gen}_\chi$ is the DM coupling used at the generator level. We remove this region in the $sA$ model to avoid violating perturbative unitarity, which can lead to an unphysical enhancement of the cross-section when the DM mass is much larger than the mediator mass \cite{Chala:2015ama,Bell:2015rdw}. The upper limit on $\sqrtgqgX$ is relatively constant across values of $\gX / \gq$, as is expected when the coupling (and hence the width) has been demonstrated to have little effect on kinematic behaviour (see sec~\ref{width_effects}), and using the assumptions of eq.~\ref{eq:sigma_propto_couplings_schan}. As the ratio increases, points in the region $\Mmed > \mX$ disappear as the initial value, $\gq = 1$, leads to a failure of the width condition. However, one could easily choose a smaller initial value of $\gq$ to recover these points, and we suggest that the limits in this region would be quite similar to those seen in the $\gX / \gq$ = 0.2 and 0.5 cases.

The constraints on the coupling strength are weaker when $\mX$ or $\Mmed$ is large ($>$100 GeV) owing to suppression of the cross-section. In this region, the constraints are expected to improve at higher centre-of-mass energies. For small DM masses with an off-shell mediator, the $\met$ distribution is softer, therefore results in this region of phase space are limited by statistical uncertainties associated with the tail-end of the distribution. This region of phase space would benefit from further optimisation of event selection in analyses aimed at the study of simplified models, as we expect to see in the upcoming Run II results.

These mono-$X$ searches are complementary to direct searches for the mediator via dijet resonances \cite{Chatrchyan:2013qha, Aad:2014aqa, Aaltonen:2008dn, Khachatryan:2015sja}. These have been used to study SiMs in, for example, \cite{An:2012va,Chala:2015ama, Zurek:tchannel}. Dijet studies search for the signature of a direct mediator decay into standard model particles, generally assuming a narrow resonance. These constraints can be stronger than mono-$X$ constraints, particularly when the width is small and when the coupling to quarks is large relative to the coupling to DM. Mono-$X$ searches however have the advantage for larger values of $\gX/\gq$ and smaller mediator masses.

We now examine channel-specific trends.

\subsubsection{Mono-jet channel}

The \monojet channel upper limits on the coupling combination $\sqrtgqgX$ for the $sV$ and $sA$ models are displayed in the left-hand column of figs.~\ref{fig:results_sVsA_rat05}-\ref{fig:results_sVsA_rat5}, for $\gX/\gq =$ 0.5, 1, 2 and 5 respectively (where the ratio of 5 is only shown for the $sV$ model, due to a lack of meaningful results in the $sA$ model). The $\gX/\gq =$ 0.2 case is shown separately in fig.~\ref{fig:results_sVsA_rat02}, as these limits are only meaningful within this channel.

As expected, the \monojet channel produces the strongest coupling limits for both $s$-channel models, which are better than those from the next-best \monoZ channel by a factor of a few. For these models, the weakest limits result for large $\mX$ or large $\Mmed$, and in fact are so weak that they are pushed into the region of invalidity where $\Gamma > \Mmed/2$. Although the acceptance is considerably higher when both $\mX$ and $\Mmed$ are large compared to low masses, the cross-section is sufficiently small so as to nullify any gain. Within the valid region ($\mX \in [1, 100]$ GeV and $\Mmed \in [1, 200]$ GeV), the limit on $\sqrtgqgX$ generally ranges from 0.1 to 0.7, with a handful of on-shell masses reaching a limit of $\sim$0.05 in the large $\gX / \gq$ case. In the large $\gX/\gq$ scenario, limits for $\mX = 1000$ GeV start to become valid; where $\sqrtgqgX$ remains constant but $\gX / \gq$ increases, the value of $\gq$ is pushed downward and so the width, which is dominated by decays to SM particles, decreases with respect to $\Mmed$.

The uncertainties on the limits for both $s$-channel models are dominated by contributions from the matching scale at acceptance-level, and generally range from $\sim5$\% to 46\%.

\subsubsection{Mono-$Z$(lep) channel}

The simplicity of the \monoZ channel relative to the \monojet channel, and the ease of signal simulation at \MG level allowed us to study a finer granularity of points in the mass phase space. The resulting limits on the sV and sA models are shown in the central column of figs.~\ref{fig:results_sVsA_rat05}-\ref{fig:results_sVsA_rat5}. While the behaviour of the limits as $\gX / \gq$ is varied is similar to that within the \monojet channel, the \monoZ limits are overall weaker by a factor of a few.

The total relative uncertainties on $\sqrtgqgX$ for the $s$-channel models are generally within 10\%, but can range up to 80\% in a few cases where $\mX$ is small; they are in general split equally between statistical and systematic contributions.

The advantage of the mono-boson channels is in the study of the $tS$ model; since this was not included in the \monojet channel the strongest limits are obtained with the \monoZ analysis, and are shown in the left-hand side of fig.~\ref{fig:results_tS}. Note that, in comparison to the $s$-channel models, the limits have weakened by a factor of 10. This is the result of an orders-of-magnitude weaker cross-section and the inability of the mediator to go on-shell in this channel. We find stronger limits for smaller $\mX$ and $\Mmed$ masses, where larger cross-sections compensate for lower acceptances at these points. Overall, the uncertainties contribute less than 10\%.

\subsubsection{Mono-$W/Z$(had) channel}

The limits on the couplings of the $sV$, $sA$ and $tS$ models, obtained within the \monoWZ channel, are shown in the right-hand column of figs.~\ref{fig:results_sVsA_rat05}-\ref{fig:results_tS}. This channel was included for comparison with the leptonic \monoZ channel in particular, but a coarser selection of masses was chosen as the limits were initially found to be somewhat weaker. Additionally, further estimates were made: a) as the kinematic behaviour is reasonably independent of the couplings, a single acceptance was found for each ($\mX$, $\Mmed$) combination and applied to each value of $\gX / \gq$, and b) complete systematic uncertainties were generated for a subset of masses and compared to those from the \monoZ channel; from this comparison the \monoZ systematic uncertainties were multiplied by 2 and then applied to the \monoWZ limits. As a result, the limits obtained in this channel are not intended to be rigorously quantitative; rather, they are used to indicate qualitatively how the channel compares.

The ATLAS \monoWZ analysis (and in particular the higher $\met$ signal region) was not optimised for a SiM interpretation, and much of the phase space produced insignificant numbers of events passing the event selection, with up to 200 thousand events generated. Generally, the limits are a factor of a few weaker than those from the \monoZ channel, which is both consistent with the limits on the EFT models studied in the ATLAS analyses, and expected following our use of a cut-and-count interpretation, rather than a shape analysis, of the \monoWZ public results.

In some cases the limits become comparable with the \monoZ channel, suggesting that more statistics and an improved treatment of systematic uncertainties would bring these closer in line with that channel.

Overall, the uncertainties from this channel lie within 5 to 50\%, most of the time being between 10 and 30\%. Generally, both statistical and systematic uncertainties contribute in a similar manner. A few points are clearly limited by the generated statistics, resulting in a statistical error of up to 90\%. Points with high $\mX$ and low $\Mmed$ tend to have larger systematic uncertainties.

\subsection{Comparison with relic density constraints}

In figs.~\ref{fig:results_sVsA_rat05}-\ref{fig:results_tS} we show lines where the constraint on the coupling corresponds to the coupling strength that would reproduce the correct DM relic density if DM is a thermal relic of the early universe. For points diagonally above and to the left of the solid purple line, the LHC constraints naively rule out the couplings leading to the correct relic density. Below and to the right of this line the relic density coupling is still allowed. In some cases the intercept does not pass through a significant number of data points surviving the quality criteria outlined in previous sections. In these cases the line is not shown.

In this scenario, the measured abundance is approximately related to the unknown self-annihilation cross-section via:
\begin{equation}
  \Omega_{\rm DM}h^2\simeq \frac{2\times2.4\times 10^{-10}\,{\rm GeV}^{-2}}{\langle\sigma v\rangle_{\rm ann}}.
  \label{simplerelic}
\end{equation}
This is used with measurements of the DM abundance by Planck, $\Omega_{\rm DM}^{\rm obs}h^2=0.1199\pm0.0027$ \cite{Ade:2013zuv}, to find $\sigv_{\rm ann}\simeq 4.0\times 10^{-9}\,{\rm GeV}^{-2}$ for thermal relic DM.
This relation is only approximately accurate, and so we use the micrOMEGAs code \cite{Belanger:2014vza} to determine the coupling strength leading to the correct relic density for each model. This technique was verified against the semi-analytic technique outlined in e.g. ref.~\cite{Busoni:2014gta}.

If the DM mass lies at the electroweak scale, the thermal relic scenario provides a natural explanation for the observed DM density. The coupling strengths leading to the correct relic density are therefore a natural benchmark with which to compare constraints from collider (and indeed direct detection) searches. However the relic density couplings should by no means be regarded as serious constraints. If DM is not produced thermally or there is an unknown effect which modifies the evolution of the density with temperature, then eq.~\ref{simplerelic} breaks down. Additionally, in the scenario where we assume DM to be a thermal relic, we ignore the possibility of there being other annihilation channels and other beyond-SM particles contributing to the DM abundance, which, if taken into account, would also invalidate eq.~\ref{simplerelic}.

\subsection{Comparison with direct detection constraints}

\begin{sidewaysfigure}
  \centering
  \begin{subfigure}[t]{0.32\textwidth}
    \centering
    \includegraphics[width=1.\textwidth]{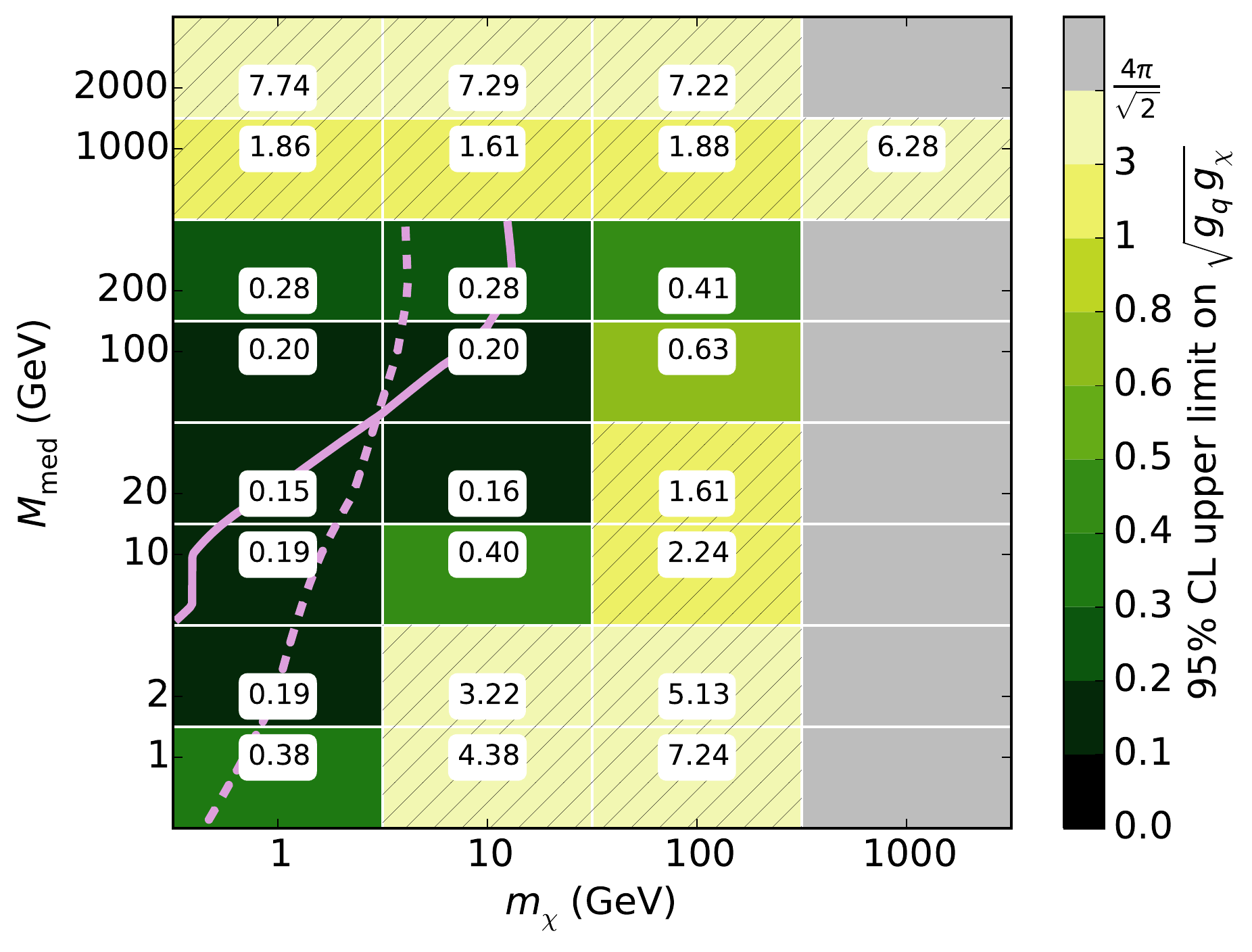}
    \caption{$sV$ model, $\gX/\gq = 0.5$, \monojet channel.}
  \end{subfigure}
  \begin{subfigure}[t]{0.32\textwidth}
    \centering
    \includegraphics[width=1.\textwidth]{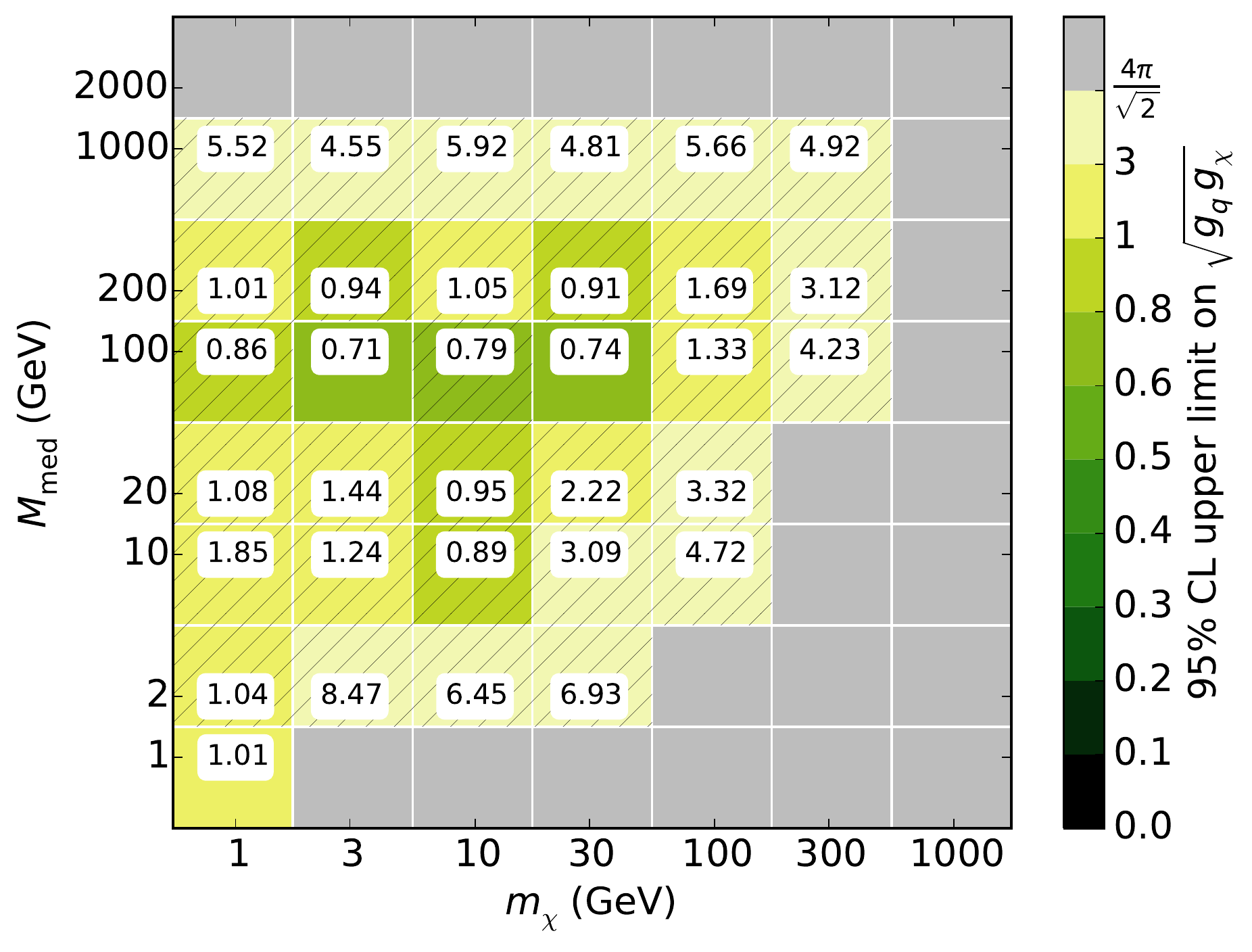}
    \caption{$sV$ model, $\gX/\gq = 0.5$, mono-$Z$ channel.}
  \end{subfigure}
  \begin{subfigure}[t]{0.32\textwidth}
    \centering
    \includegraphics[width=1.\textwidth]{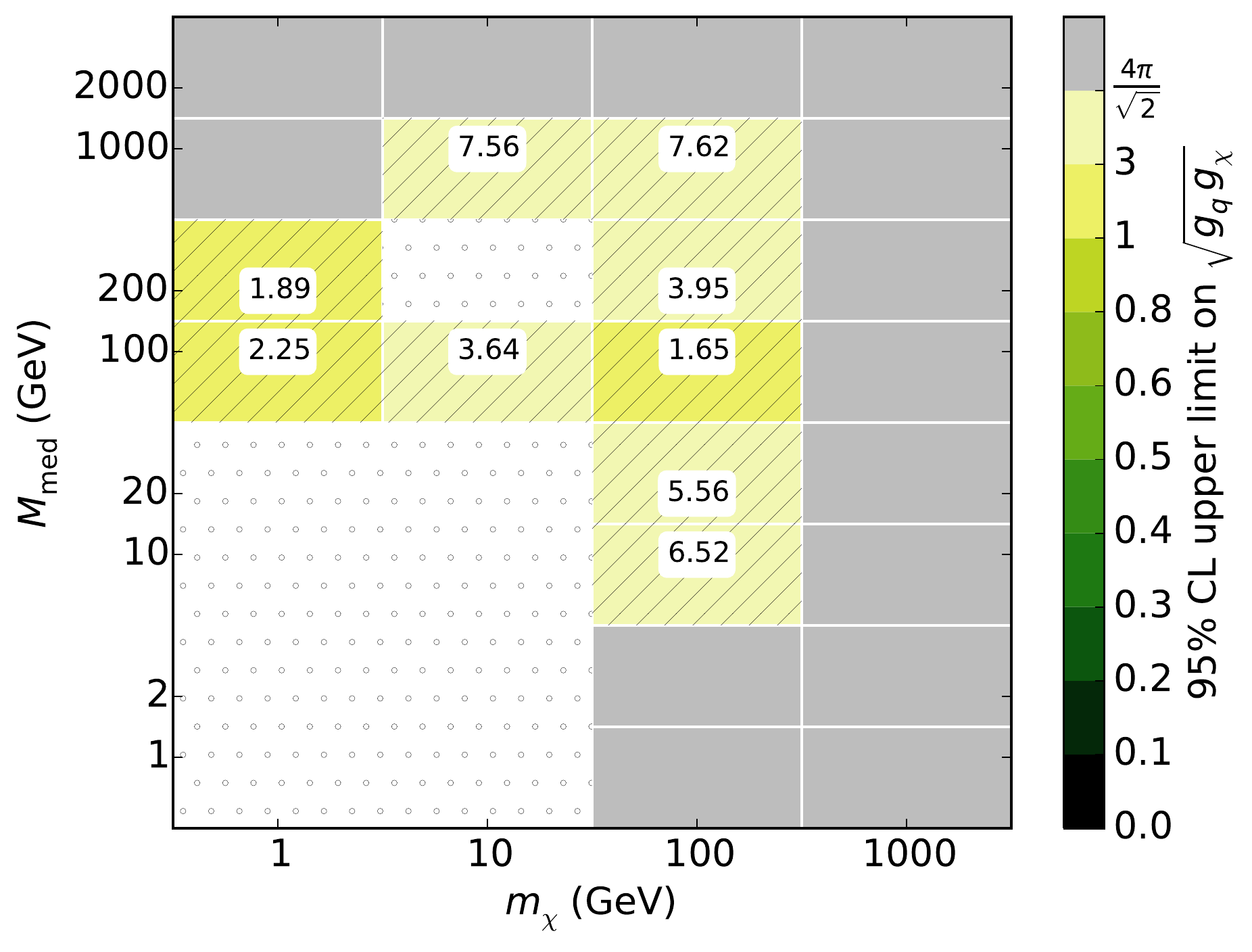}
    \caption{$sV$ model, $\gX/\gq = 0.5$, mono-$W/Z$ channel.}
    \vspace{0.75cm}
  \end{subfigure}
  \begin{subfigure}[t]{0.32\textwidth}
    \centering
    \includegraphics[width=1.\textwidth]{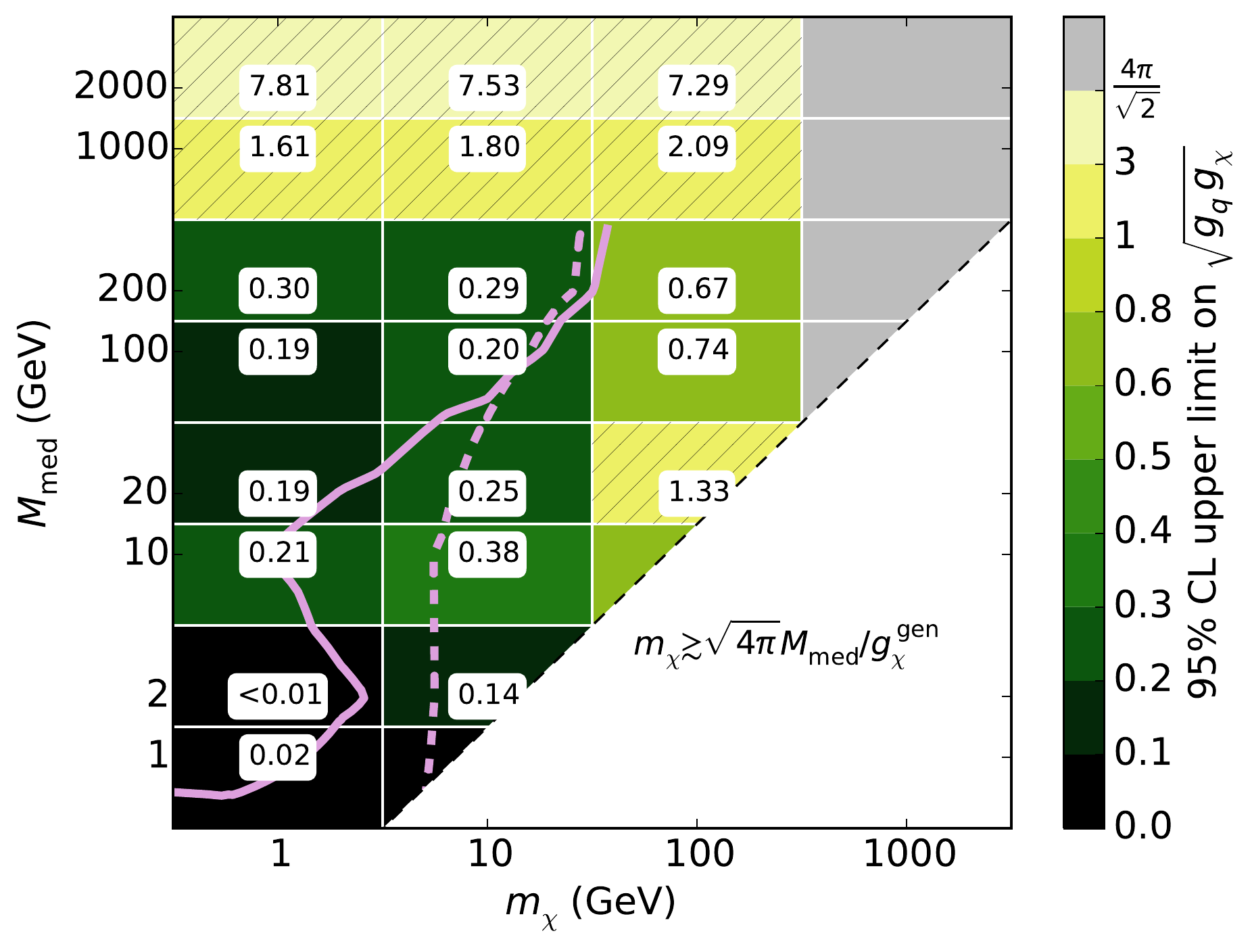}
    \caption{$sA$ model, $\gX/\gq = 0.5$, \monojet channel.}
  \end{subfigure}
  \begin{subfigure}[t]{0.32\textwidth}
    \centering
    \includegraphics[width=1.\textwidth]{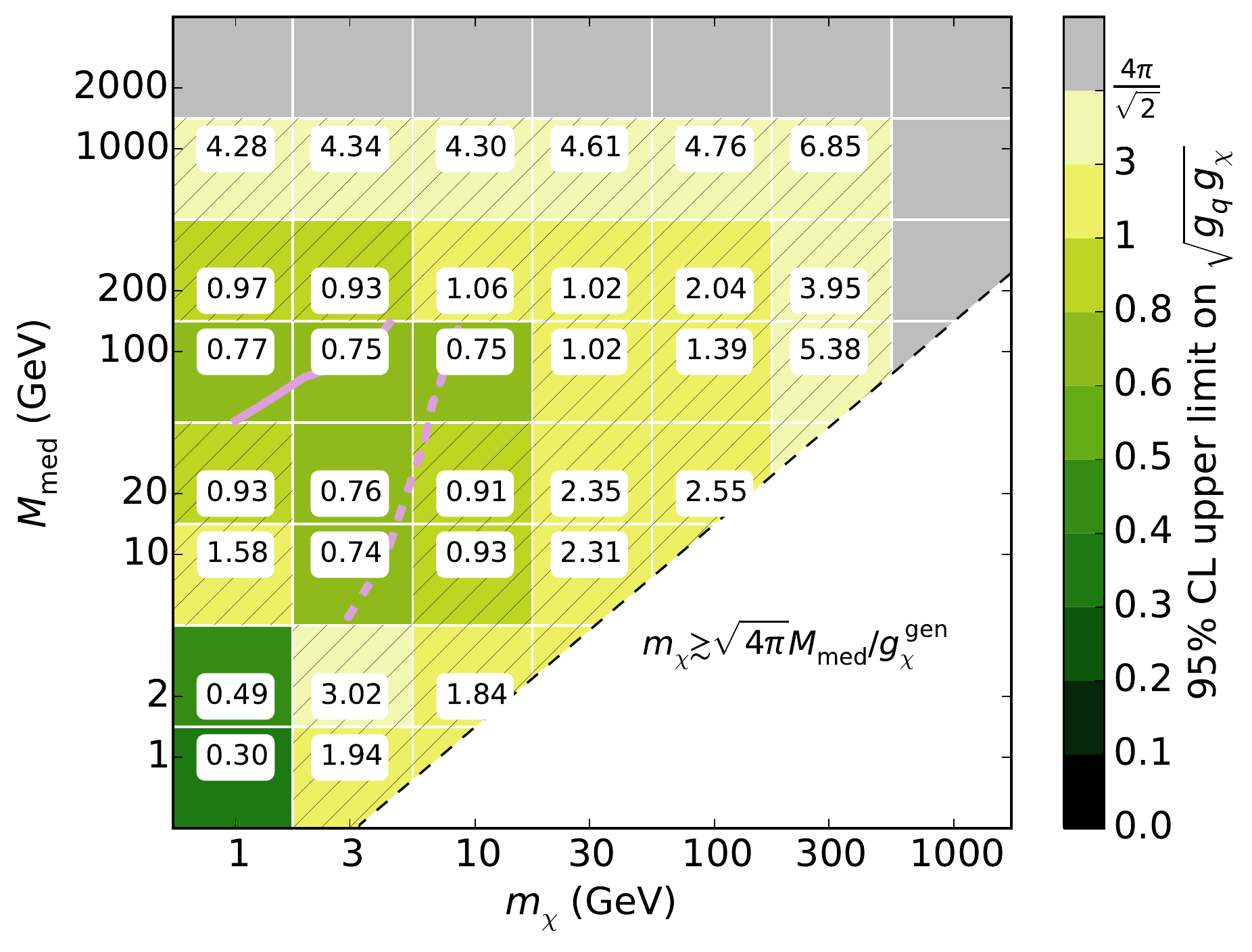}
    \caption{$sA$ model, $\gX/\gq = 0.5$, mono-$Z$ channel.}
  \end{subfigure}
  \begin{subfigure}[t]{0.32\textwidth}
    \centering
    \includegraphics[width=1.\textwidth]{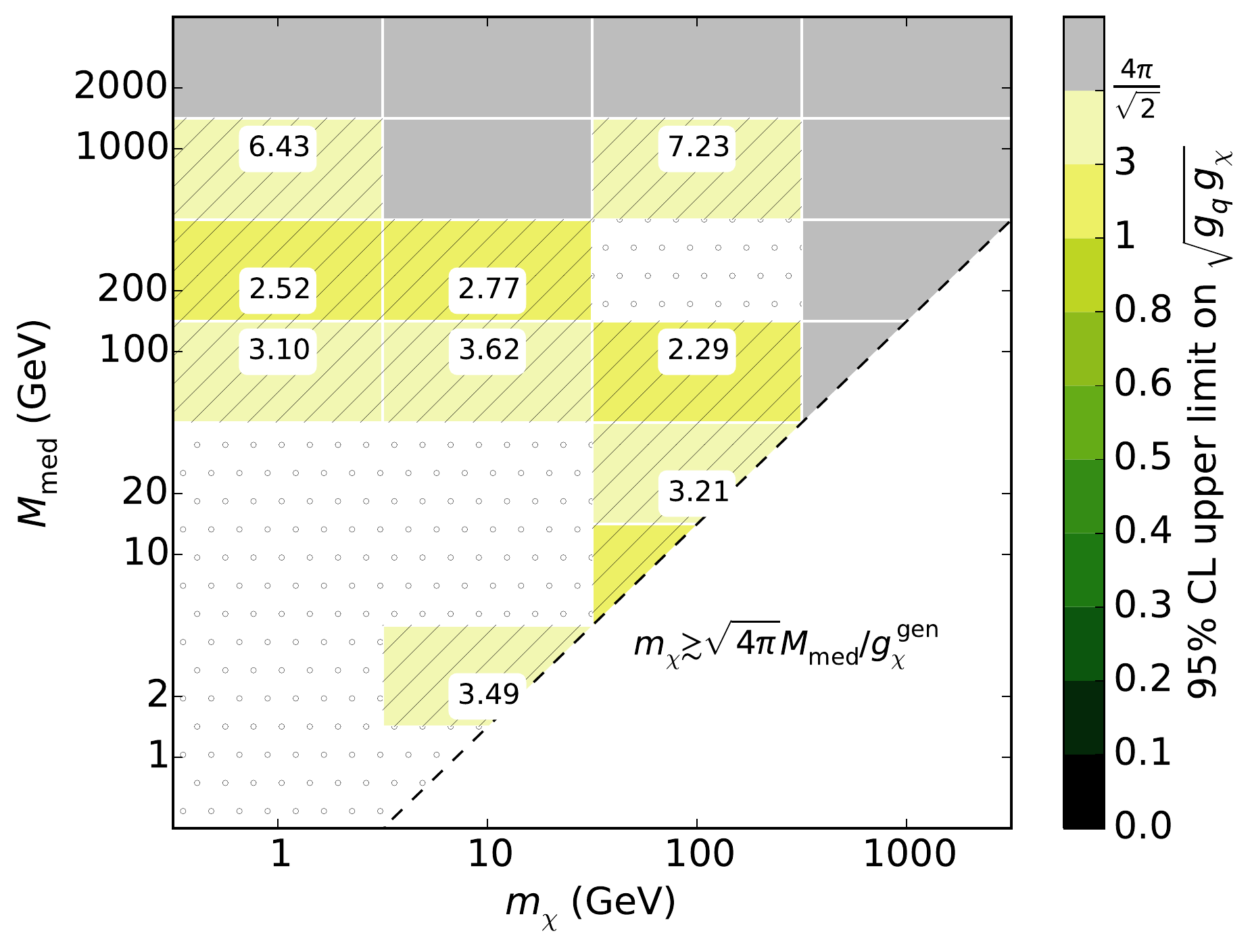}
    \caption{$sA$ model, $\gX/\gq = 0.5$, mono-$W/Z$ channel.}
  \end{subfigure}
  \caption{Upper limits on the coupling for the $s$-channel models in the \monojet (left), \monoZ (centre) and \monoWZ (right) channels, for $\gX / \gq$ = 0.5. The grey region represents the phase space where no meaningful limit was obtained. The hatched region represents a limit which leads to a width greater than $\Mmed / 2$, so the validity of the calculation begins to fail. The dotted region represents phase space where insufficient statistics were available. The purple dashed line shows the threshold where direct detection constraints become stronger than the \monoX constraint. Similarly the solid purple line shows where relic density constraints become stronger than \monoX constraints. See the text for further details.}
  \label{fig:results_sVsA_rat05}
\end{sidewaysfigure}

\begin{sidewaysfigure}
  \centering
  \begin{subfigure}[t]{0.32\textwidth}
    \centering
    \includegraphics[width=1.\textwidth]{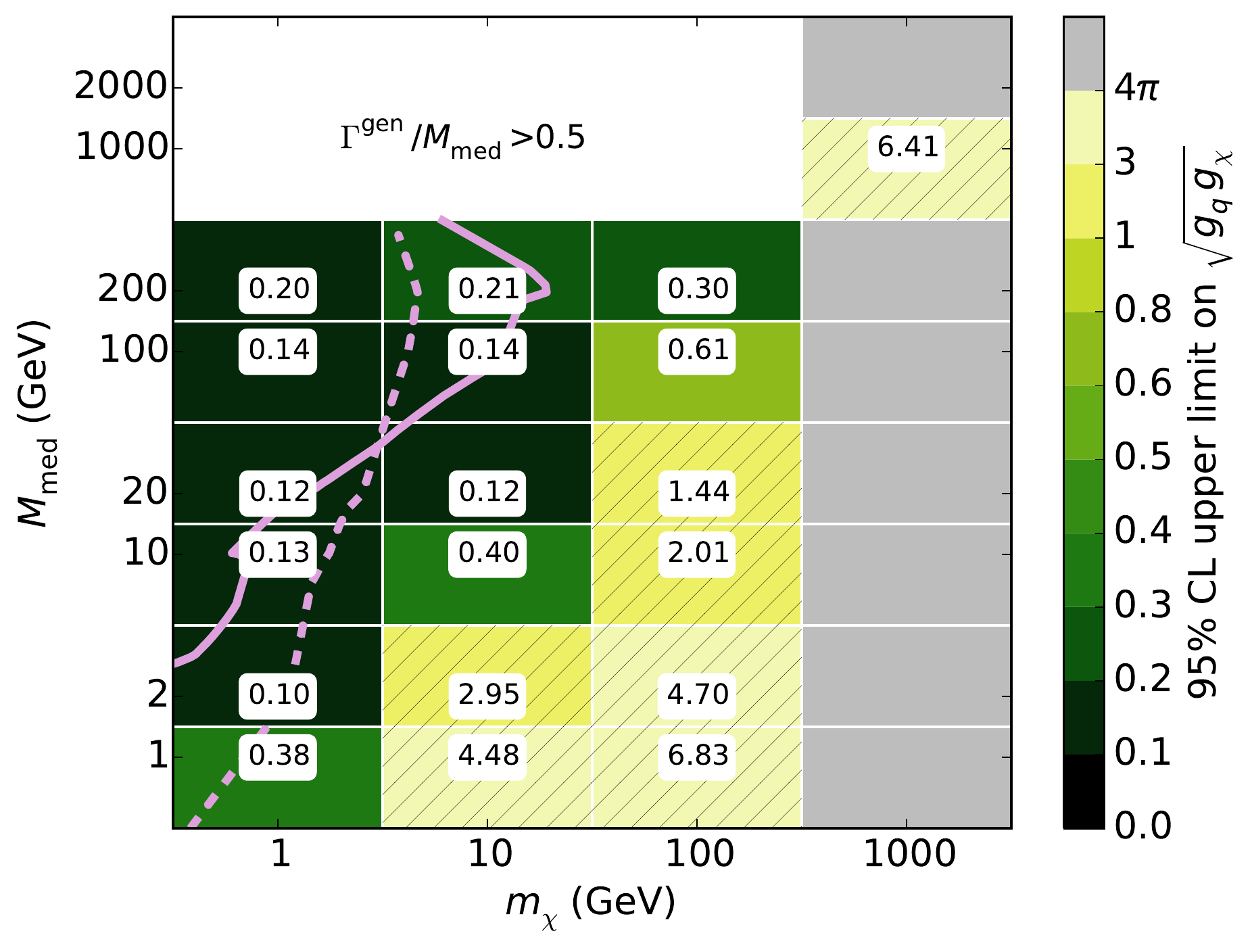}
    \caption{$sV$ model, $\gX/\gq = 1$, \monojet channel.}
  \end{subfigure}
  \begin{subfigure}[t]{0.32\textwidth}
    \centering
    \includegraphics[width=1.\textwidth]{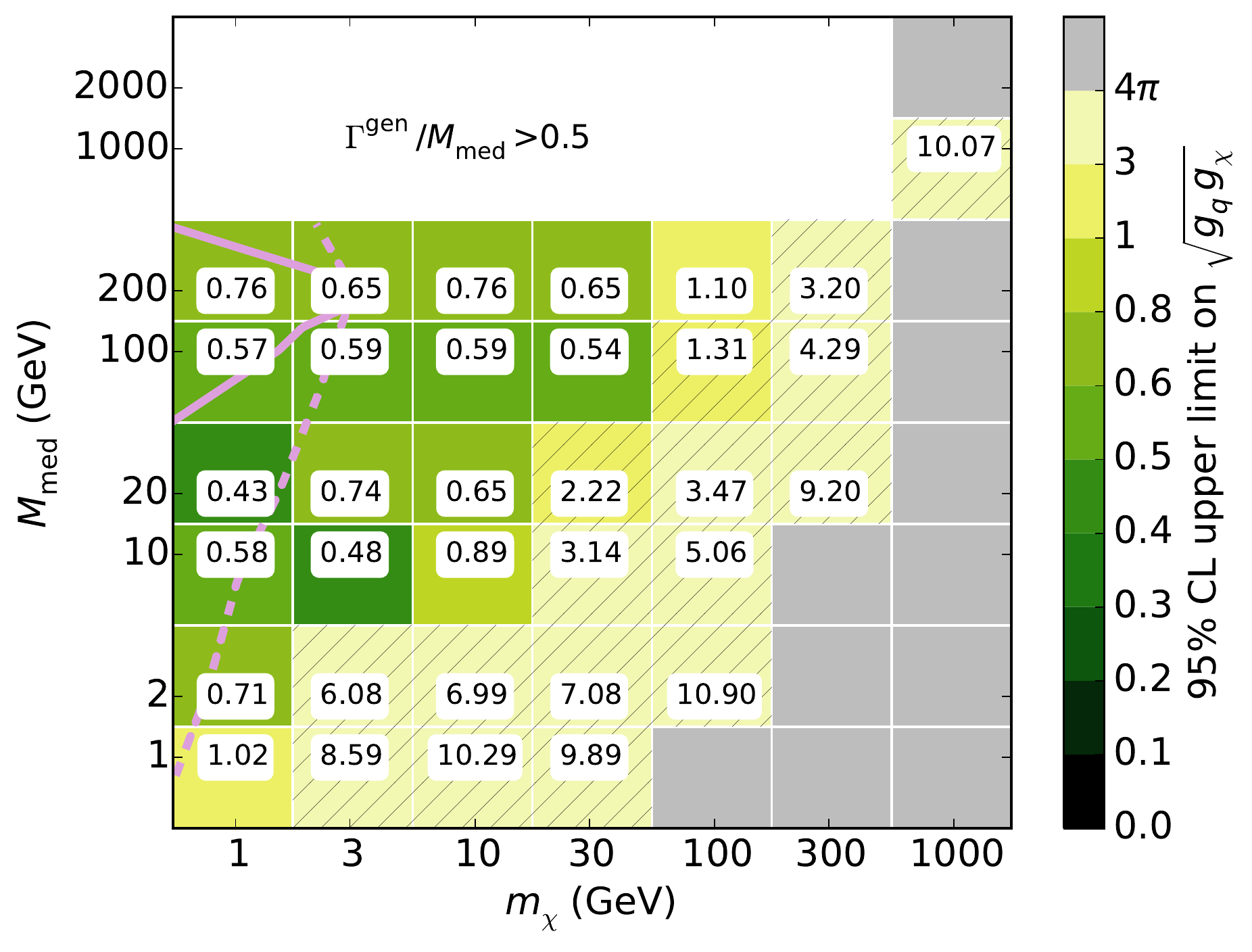}
    \caption{$sV$ model, $\gX/\gq = 1$, mono-$Z$ channel.}
  \end{subfigure}
  \begin{subfigure}[t]{0.32\textwidth}
    \centering
    \includegraphics[width=1.\textwidth]{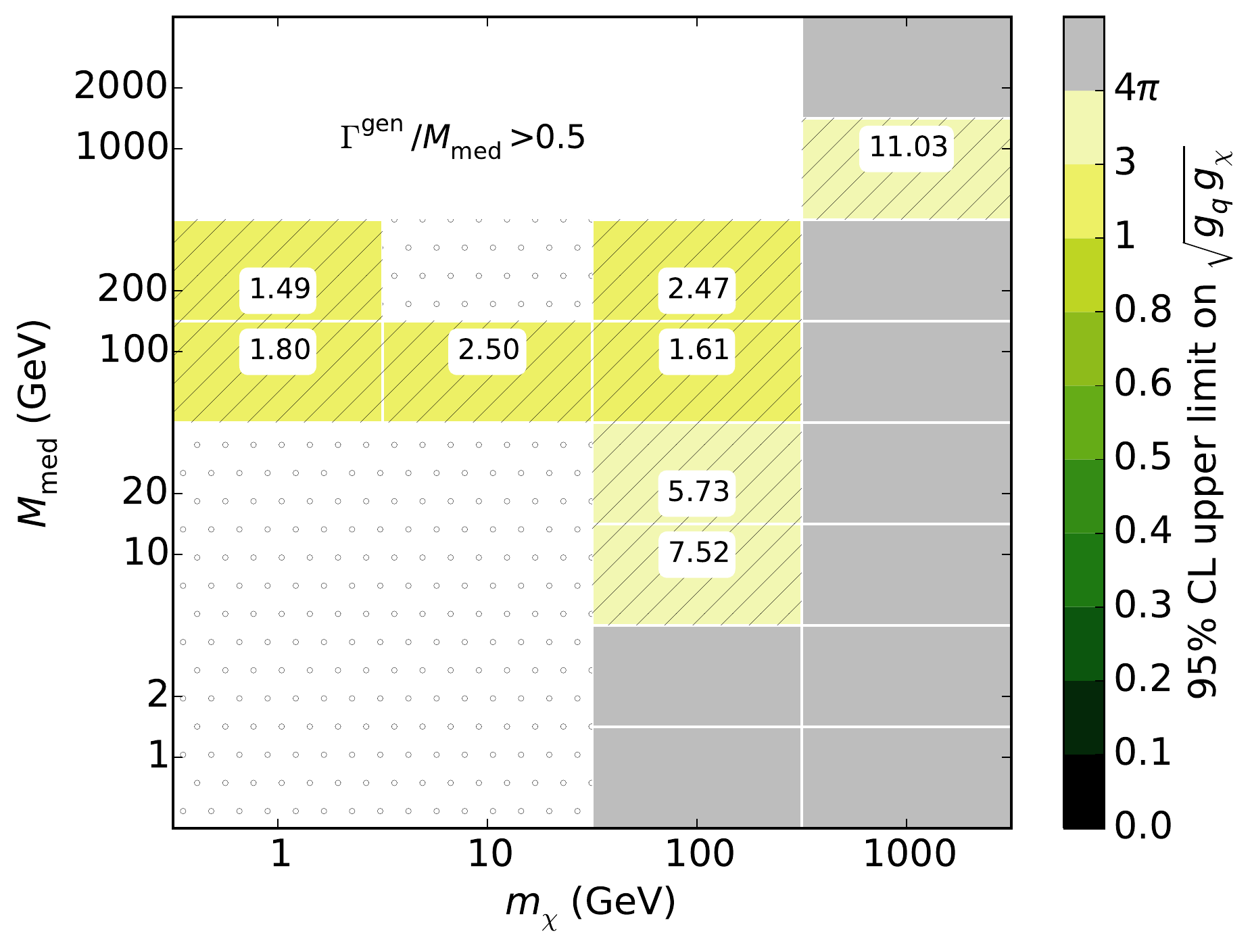}
    \caption{$sV$ model, $\gX/\gq = 1$, mono-$W/Z$ channel.}
    \vspace{0.75cm}
  \end{subfigure}
  \begin{subfigure}[t]{0.32\textwidth}
    \centering
    \includegraphics[width=1.\textwidth]{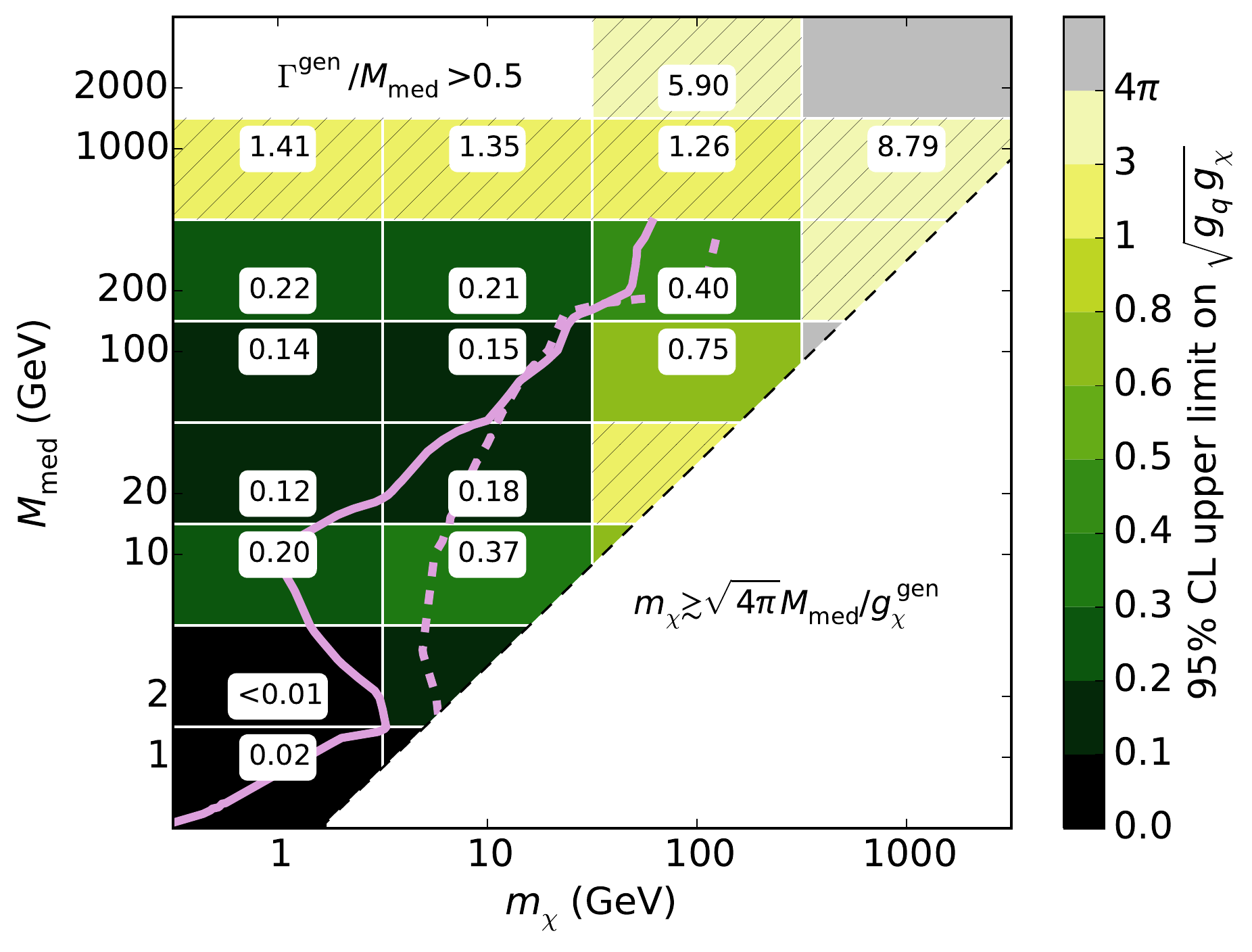}
    \caption{$sA$ model, $\gX/\gq = 1$, \monojet channel.}
  \end{subfigure}
  \begin{subfigure}[t]{0.32\textwidth}
    \centering
    \includegraphics[width=1.\textwidth]{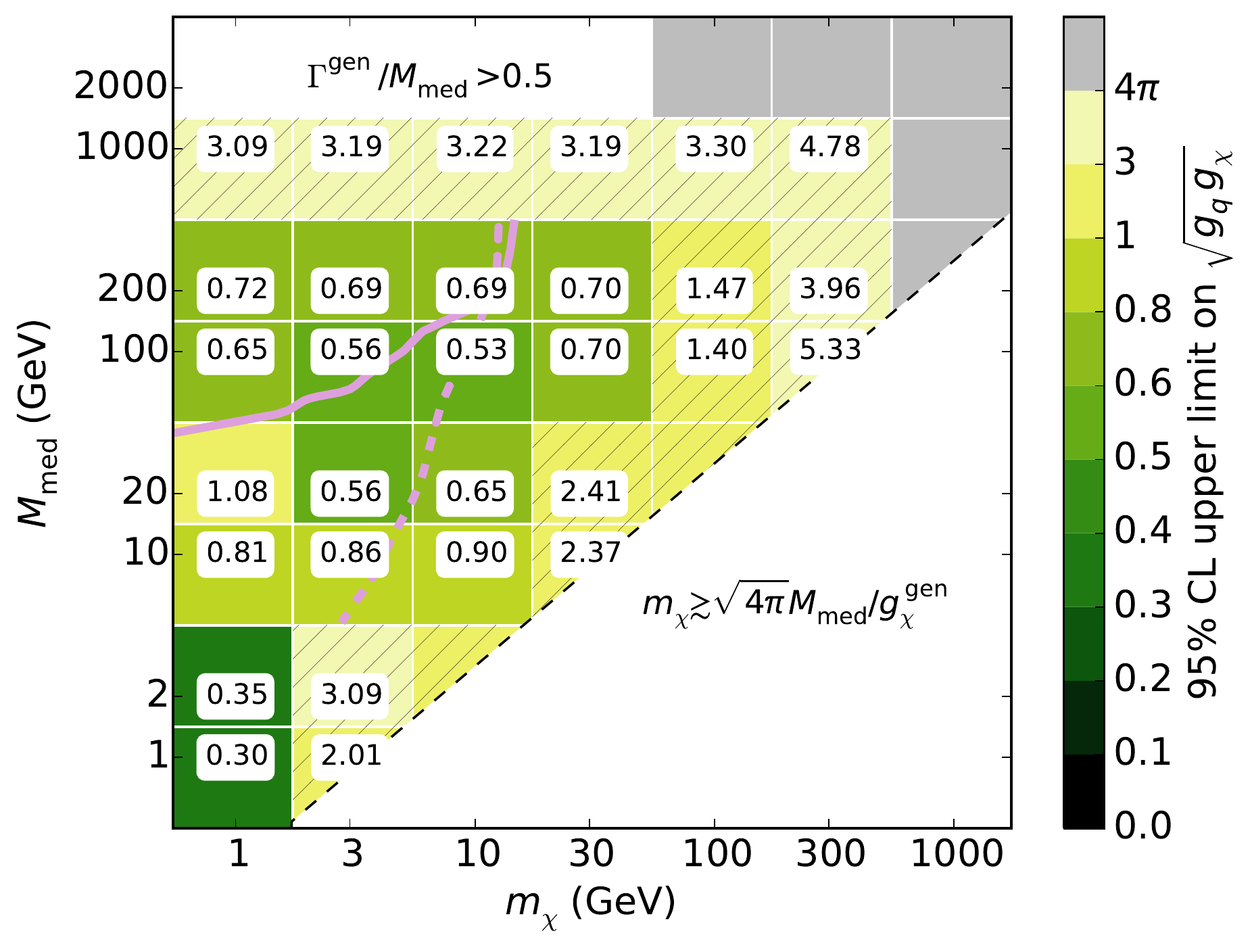}
    \caption{$sA$ model, $\gX/\gq = 1$, mono-$Z$ channel.}
  \end{subfigure}
  \begin{subfigure}[t]{0.32\textwidth}
    \centering
    \includegraphics[width=1.\textwidth]{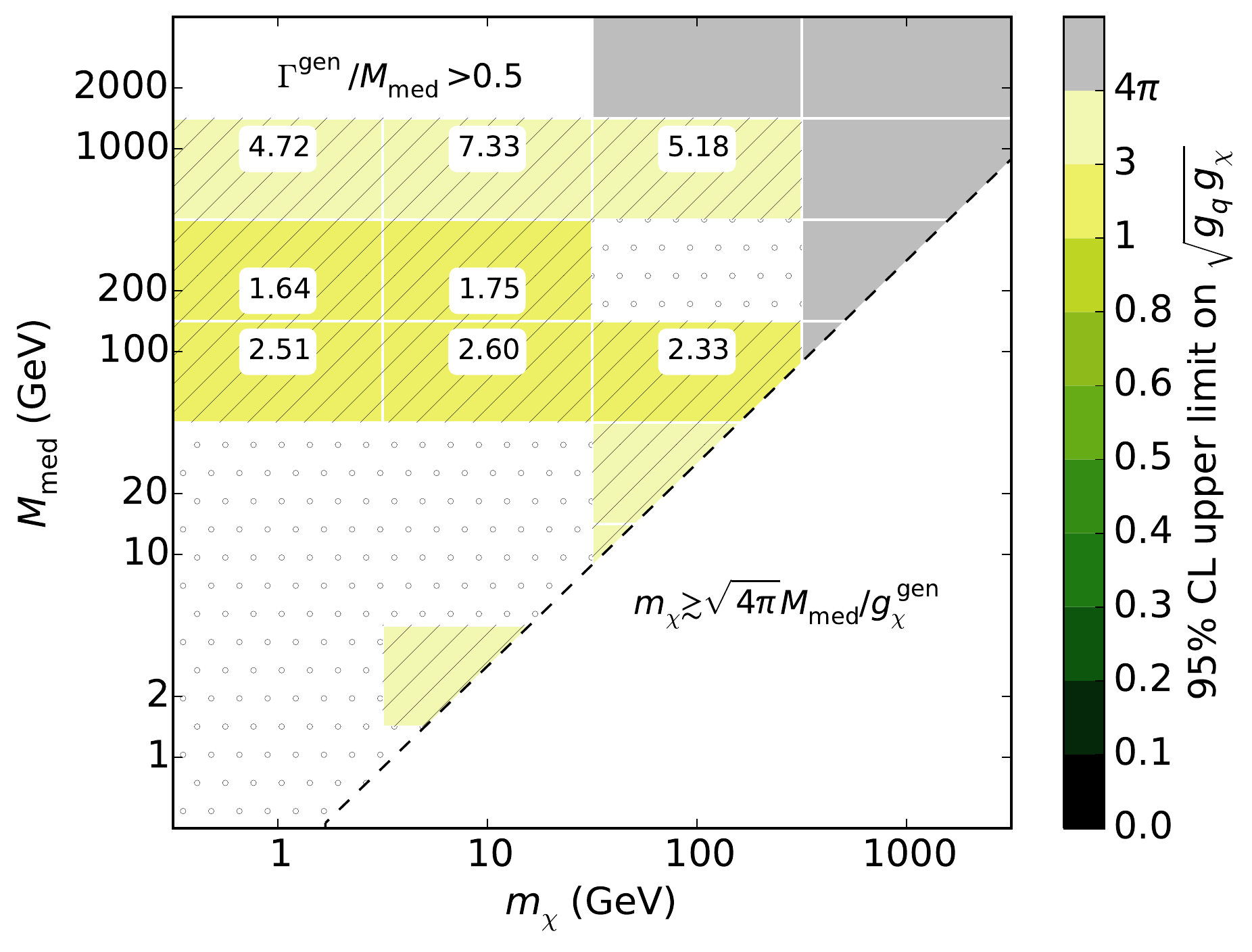}
    \caption{$sA$ model, $\gX/\gq = 1$, mono-$W/Z$ channel.}
  \end{subfigure}
  \caption{Upper limits on the couplings for the $s$-channel models in the \monojet (left), \monoZ (centre) and \monoWZ (right) channels, for $\gX / \gq$ = 1. Refer to fig.~\ref{fig:results_sVsA_rat05} for details.}
  \label{fig:results_sVsA_rat1}
\end{sidewaysfigure}

\begin{sidewaysfigure}
  \centering
  \begin{subfigure}[t]{0.32\textwidth}
    \centering
    \includegraphics[width=1.\textwidth]{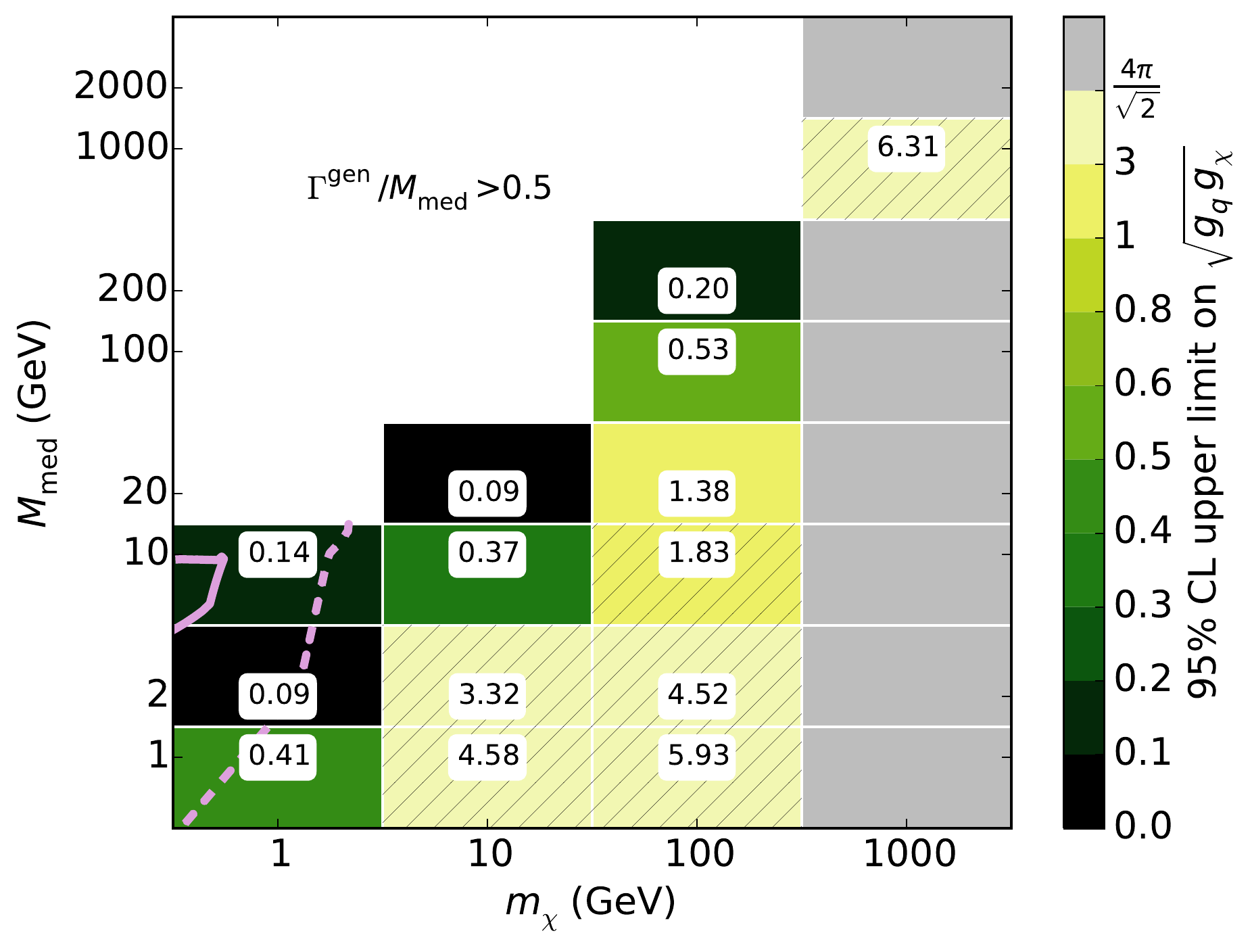}
    \caption{$sV$ model, $\gX/\gq = 2$, \monojet channel.}
  \end{subfigure}
  \begin{subfigure}[t]{0.32\textwidth}
    \centering
    \includegraphics[width=1.\textwidth]{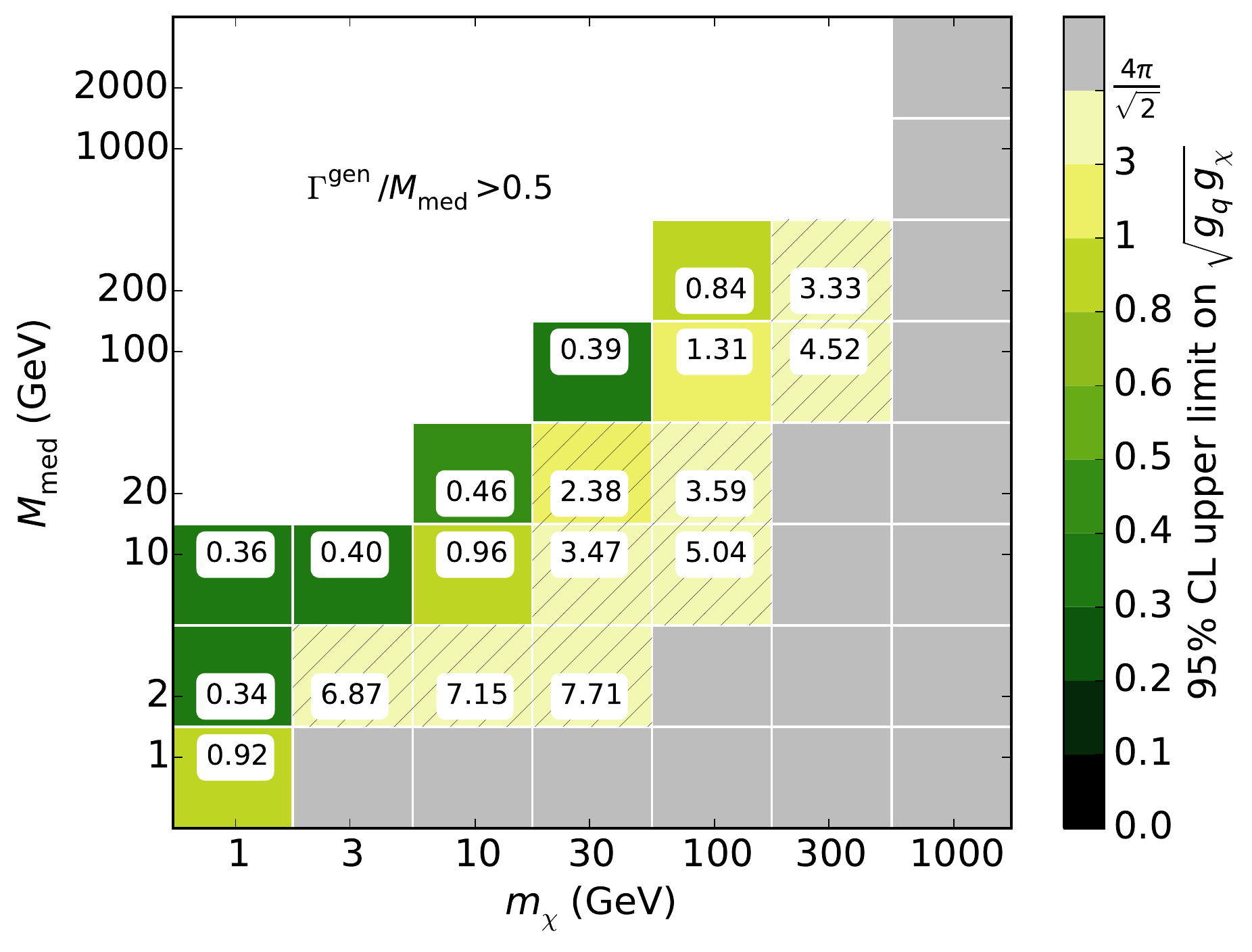}
    \caption{$sV$ model, $\gX/\gq = 2$, mono-$Z$ channel.}
  \end{subfigure}
  \begin{subfigure}[t]{0.32\textwidth}
    \centering
    \includegraphics[width=1.\textwidth]{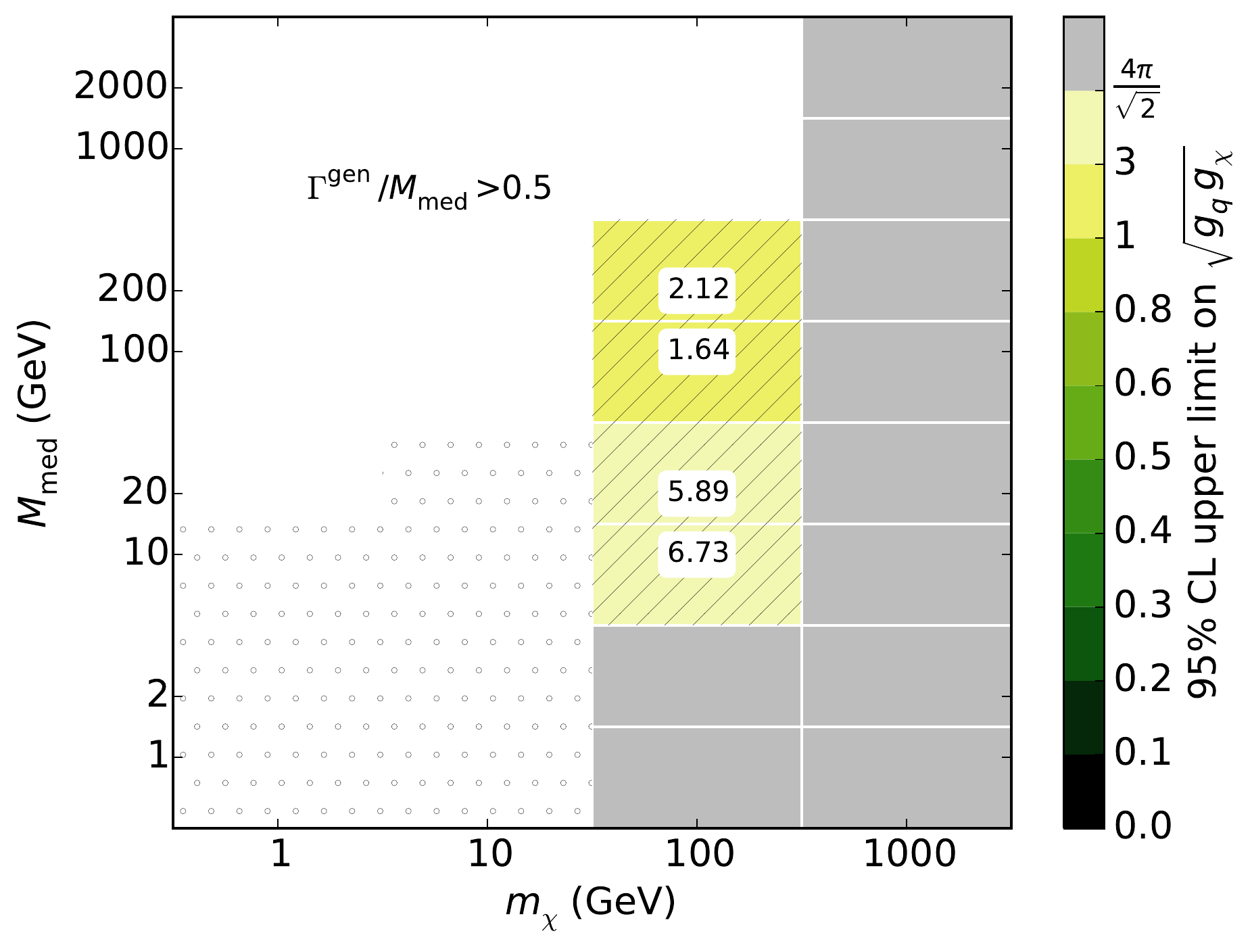}
    \caption{$sV$ model, $\gX/\gq = 2$, mono-$W/Z$ channel.}
    \vspace{0.75cm}
  \end{subfigure}
  \begin{subfigure}[t]{0.32\textwidth}
    \centering
    \includegraphics[width=1.\textwidth]{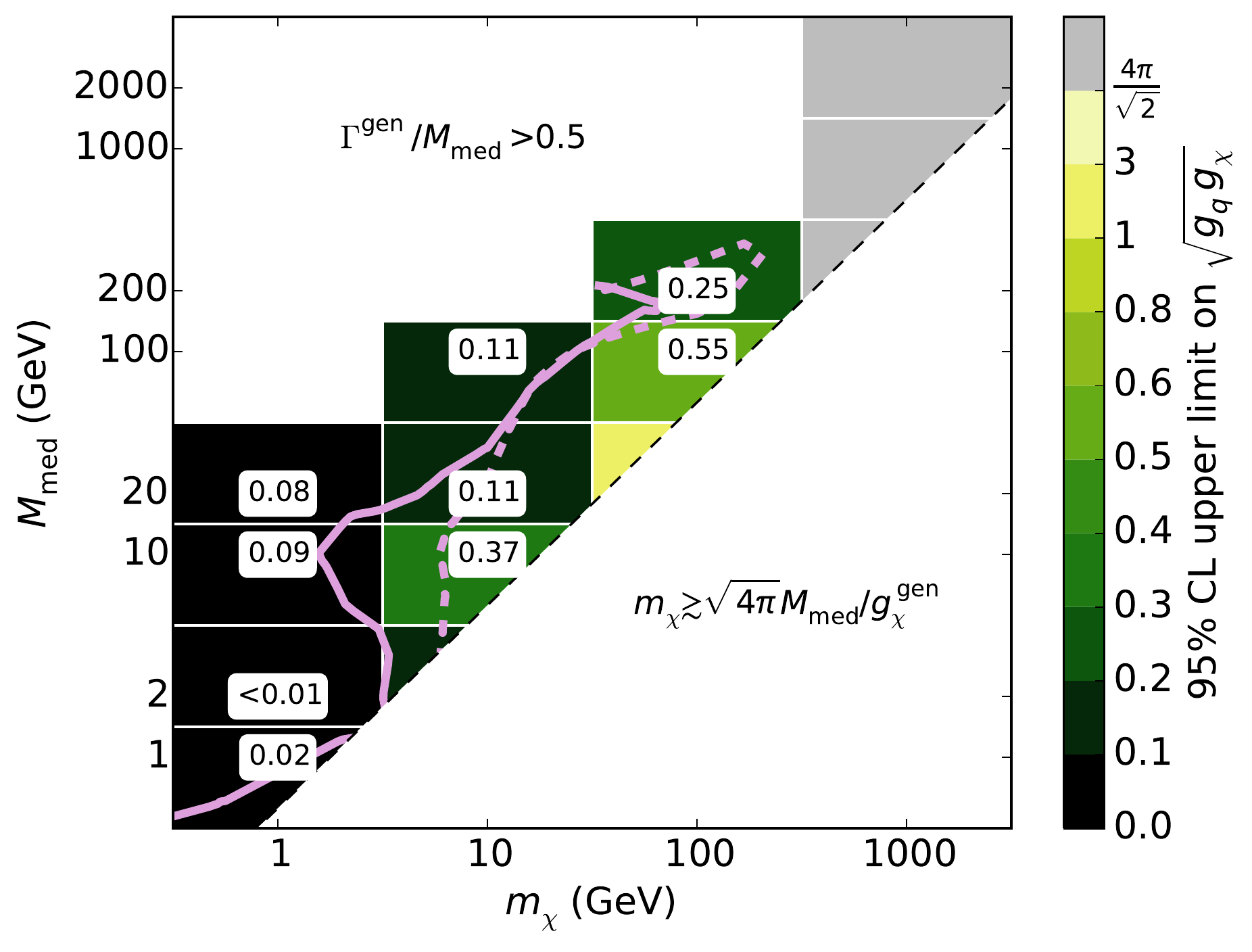}
    \caption{$sA$ model, $\gX/\gq = 2$, \monojet channel.}
  \end{subfigure}
  \begin{subfigure}[t]{0.32\textwidth}
    \centering
    \includegraphics[width=1.\textwidth]{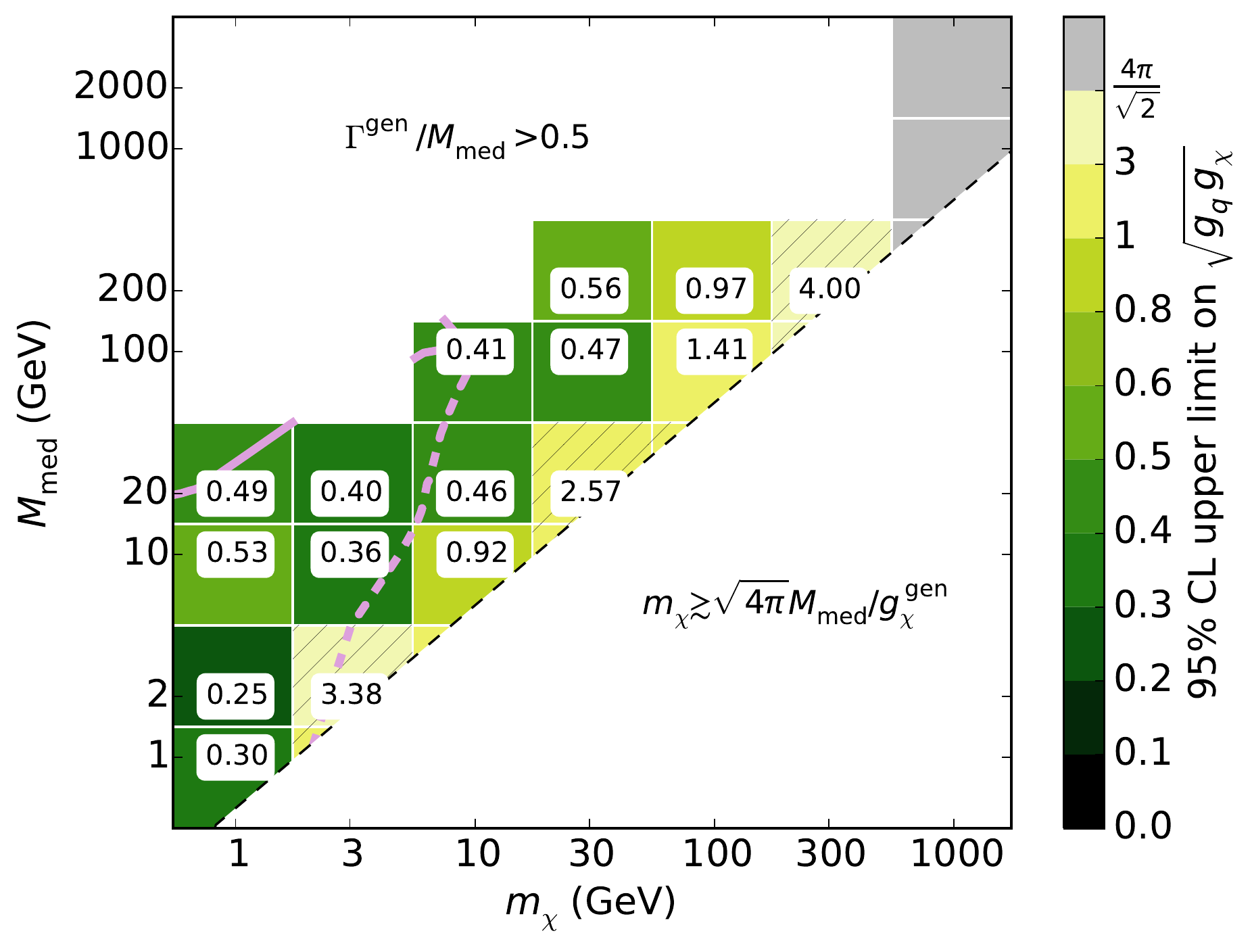}
    \caption{$sA$ model, $\gX/\gq = 2$, mono-$Z$ channel.}
  \end{subfigure}
  \begin{subfigure}[t]{0.32\textwidth}
    \centering
    \includegraphics[width=1.\textwidth]{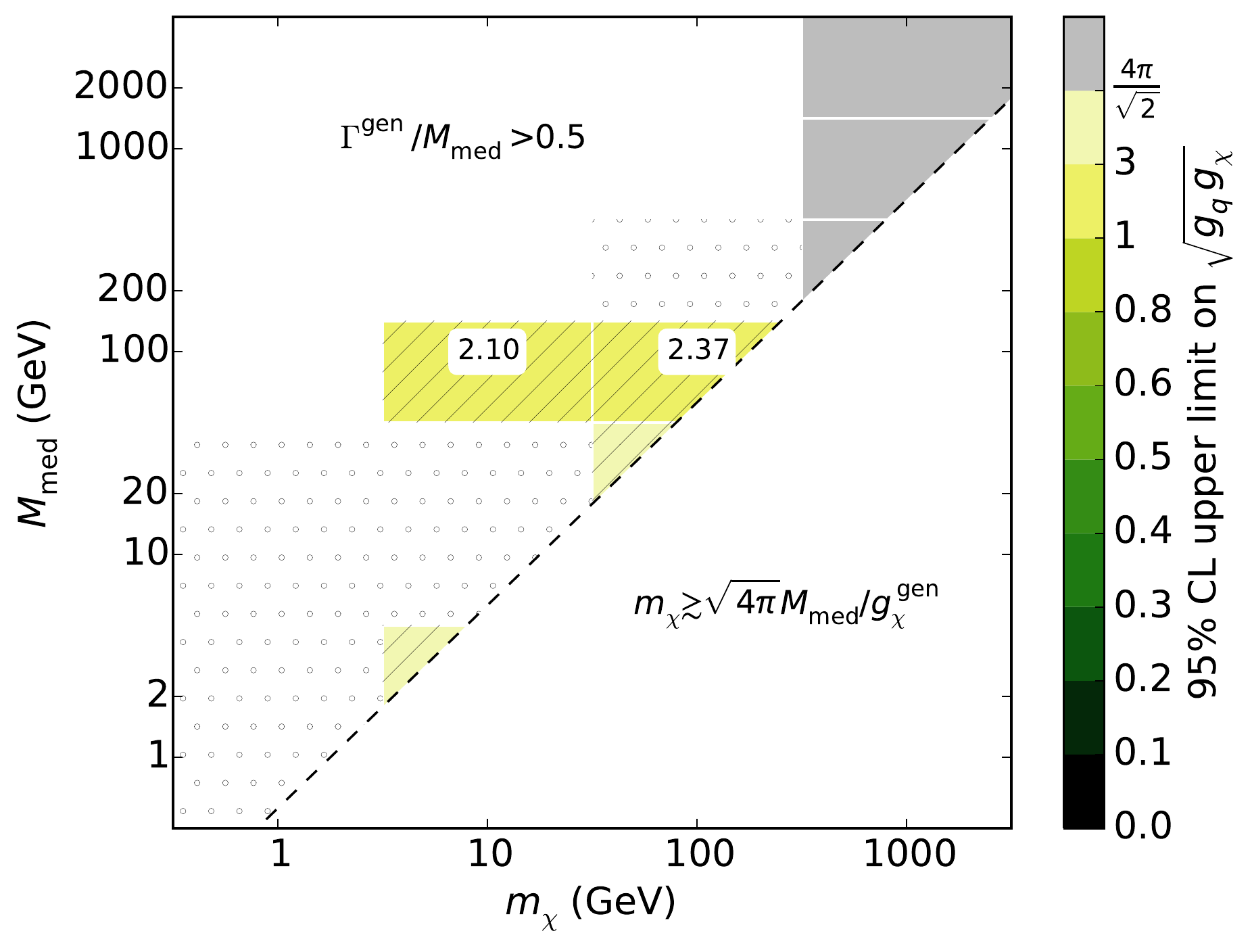}
    \caption{$sA$ model, $\gX/\gq = 2$, mono-$W/Z$ channel.}
  \end{subfigure}
  \caption{Upper limits on the coupling for the $s$-channel models in the \monojet (left), \monoZ (centre) and \monoWZ (right) channels, for $\gX / \gq$ = 2. Refer to fig.~\ref{fig:results_sVsA_rat05} for details.}
  \label{fig:results_sVsA_rat2}
\end{sidewaysfigure}

\begin{sidewaysfigure}
  \centering
  \begin{subfigure}[t]{0.32\textwidth}
    \centering
    \includegraphics[width=1.\textwidth]{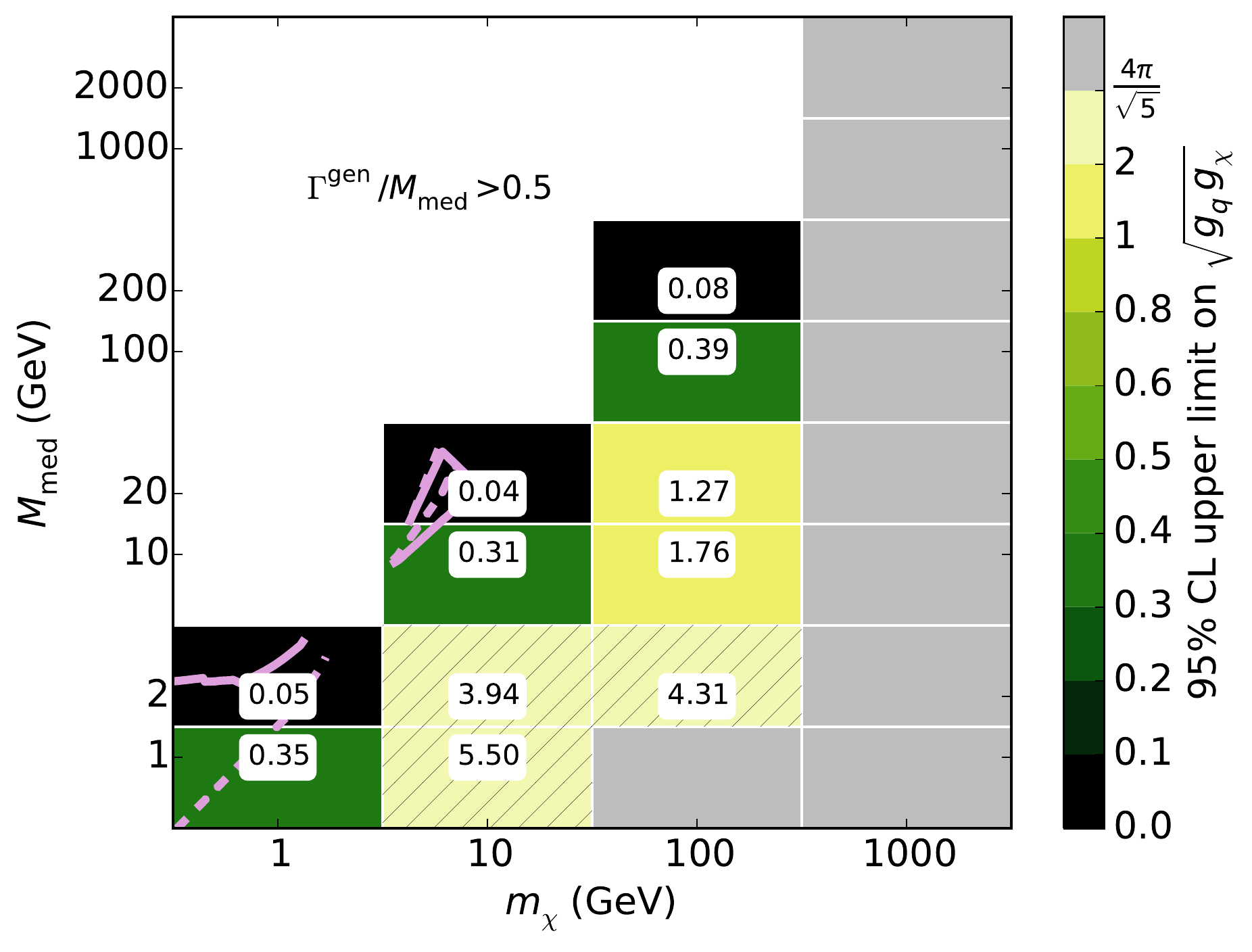}
    \caption{$sV$ model, $\gX/\gq = 5$, \monojet channel.}
  \end{subfigure}
  \begin{subfigure}[t]{0.32\textwidth}
    \centering
    \includegraphics[width=1.\textwidth]{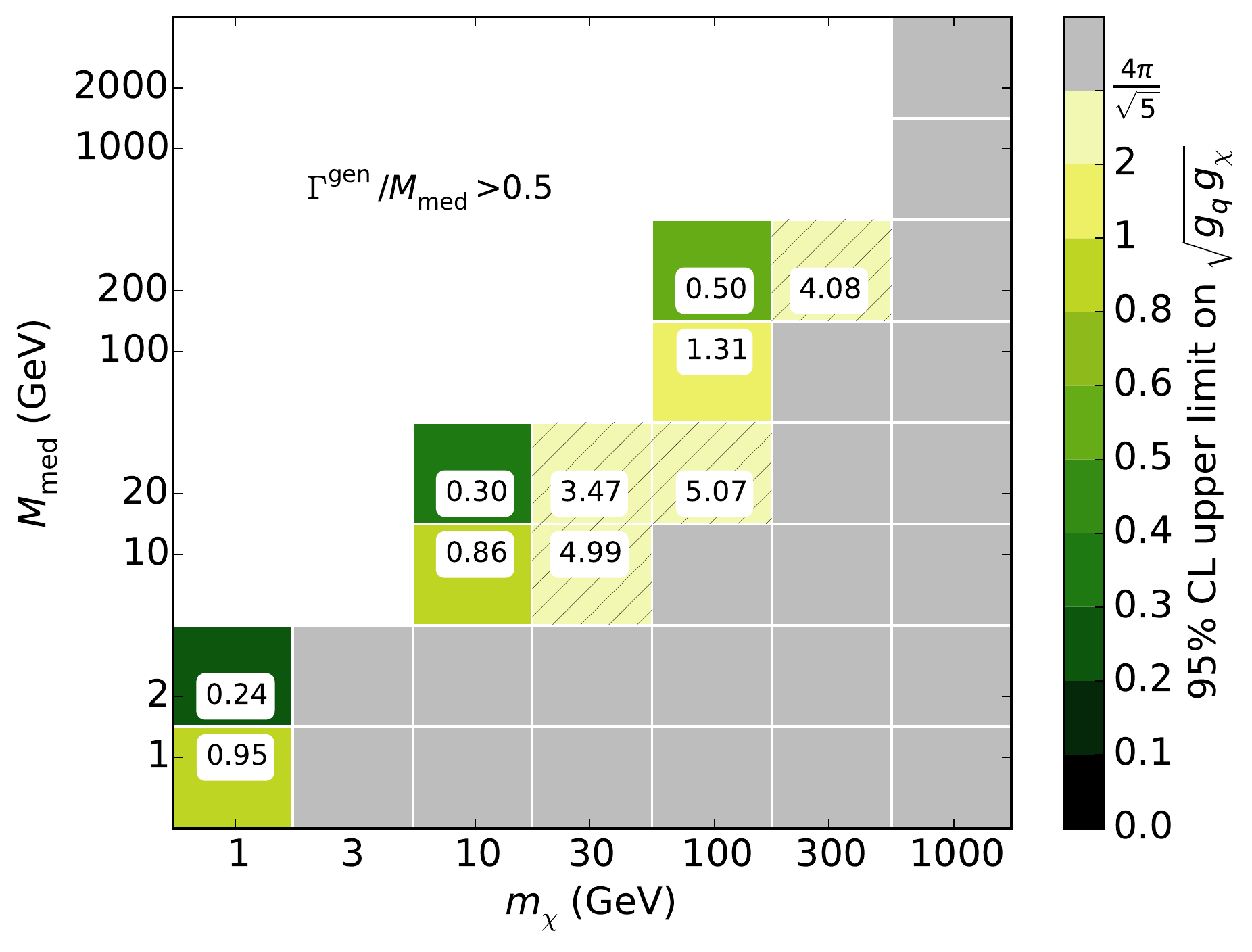}
    \caption{$sV$ model, $\gX/\gq = 5$, mono-$Z$ channel.}
  \end{subfigure}
  \begin{subfigure}[t]{0.32\textwidth}
    \centering
    \includegraphics[width=1.\textwidth]{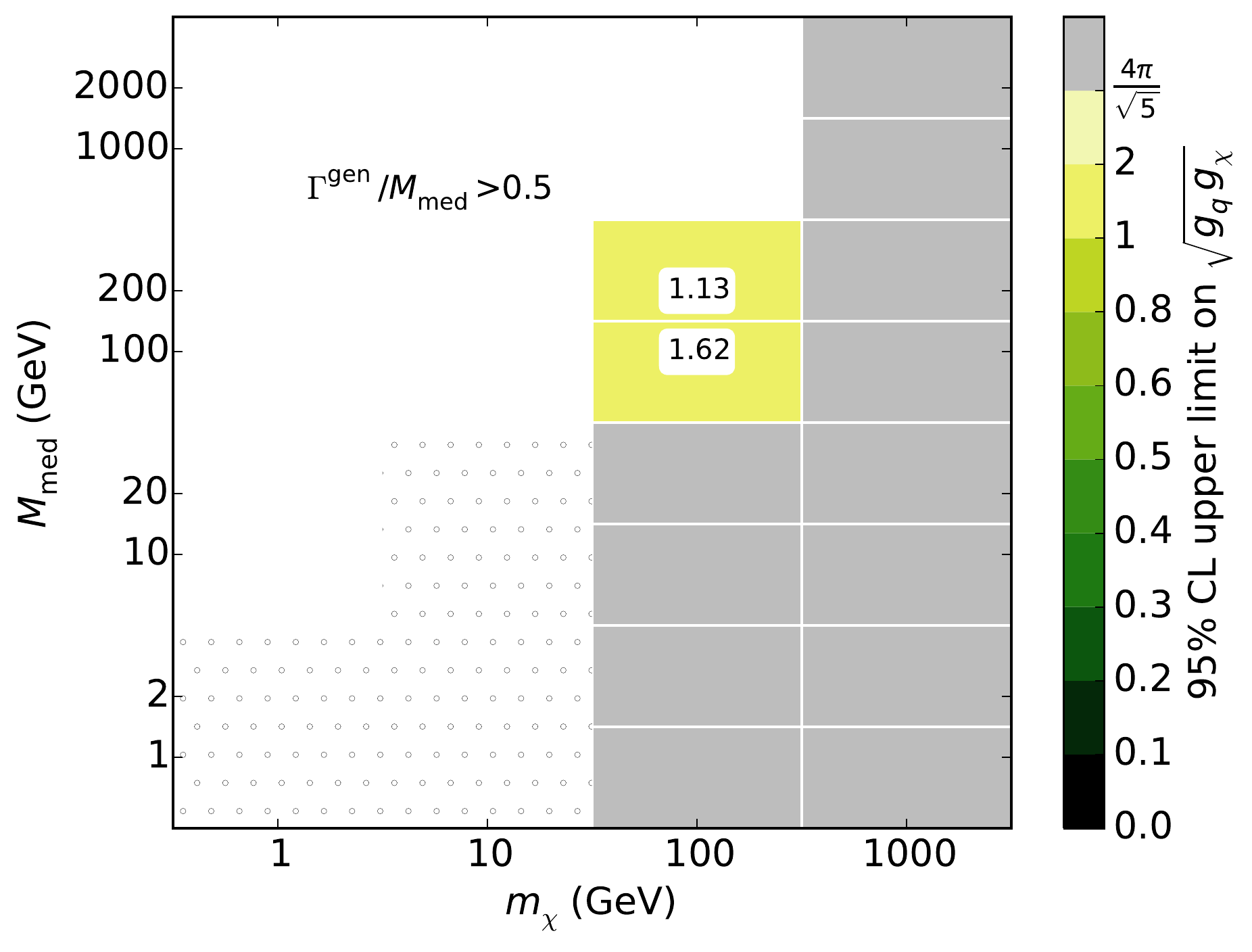}
    \caption{$sV$ model, $\gX/\gq = 5$, mono-$W/Z$ channel.}
    \vspace{0.75cm}
  \end{subfigure}
  \caption{Upper limits on the coupling for the $sV$ model in the \monojet (left), \monoZ (centre) and \monoWZ (right) channels, for $\gX / \gq$ = 5. Refer to fig.~\ref{fig:results_sVsA_rat05} for details.}
  \label{fig:results_sVsA_rat5}
\end{sidewaysfigure}

\begin{figure}
  \centering
  \begin{subfigure}[t]{0.495\textwidth}
    \centering
    \includegraphics[width=1.\textwidth]{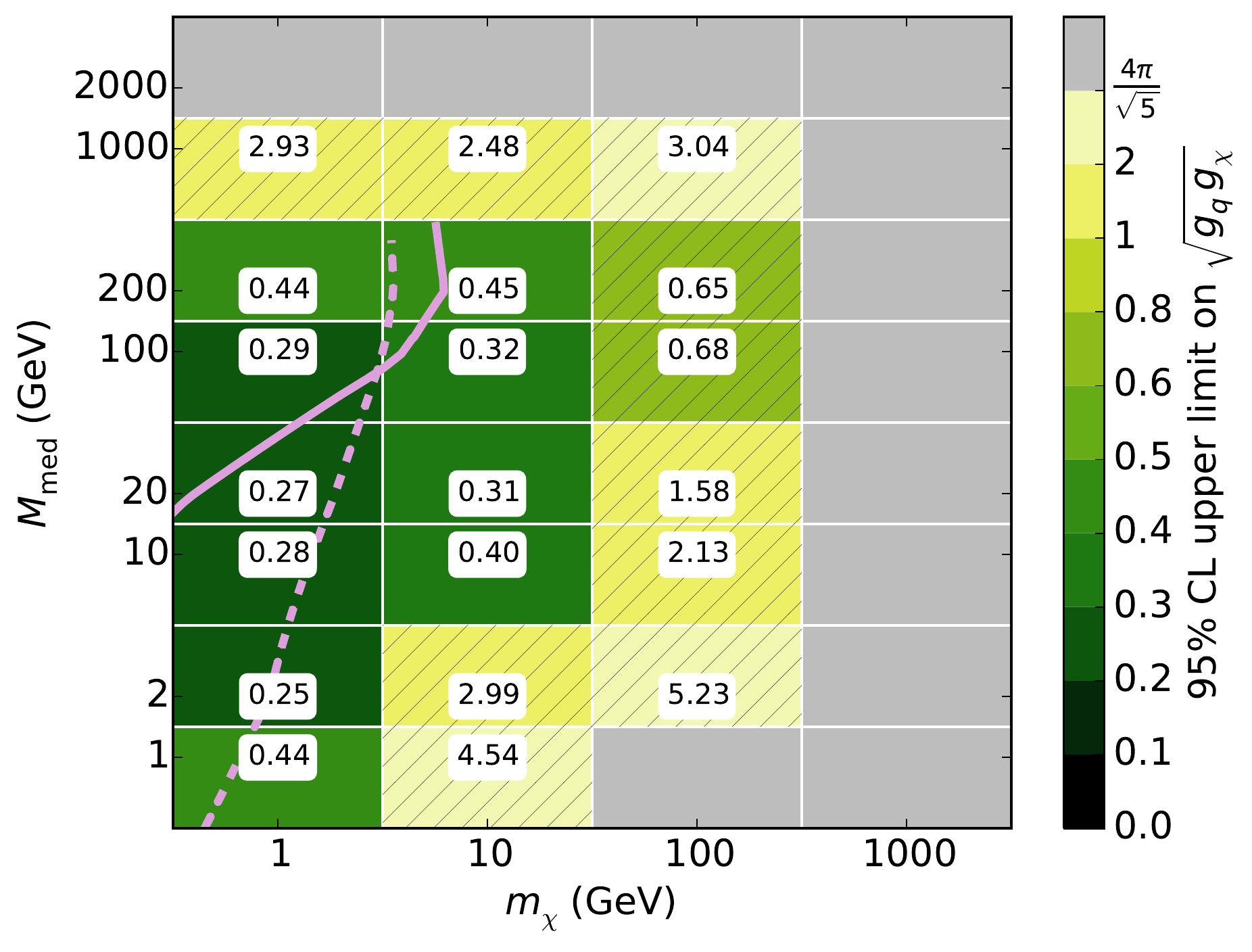}
    \caption{$sV$ model, $\gX/\gq = 0.2$, \monojet channel.}
  \end{subfigure}
  \begin{subfigure}[t]{0.495\textwidth}
    \centering
    \includegraphics[width=1.\textwidth]{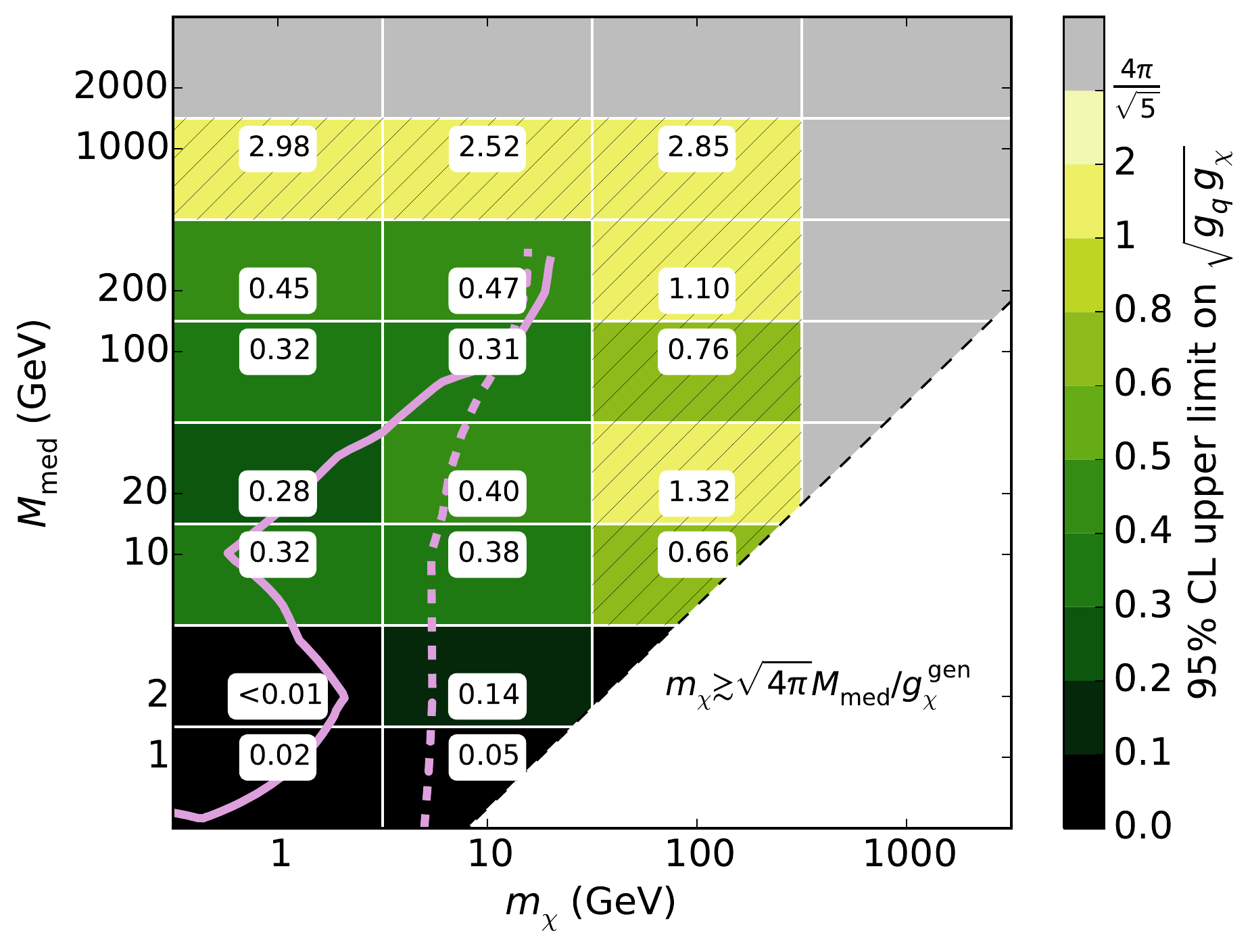}
    \caption{$sA$ model, $\gX/\gq = 0.2$, \monojet channel.}
  \end{subfigure}
  \caption{Upper limits on the coupling for the $s$-channel models in the \monojet channel, for $\gX / \gq$ = 0.2. Refer to fig.~\ref{fig:results_sVsA_rat05} for details.}
  \label{fig:results_sVsA_rat02}
\end{figure}

\begin{figure}
  \centering
  \begin{subfigure}[t]{0.495\textwidth}
    \centering
    \includegraphics[width=1.\textwidth]{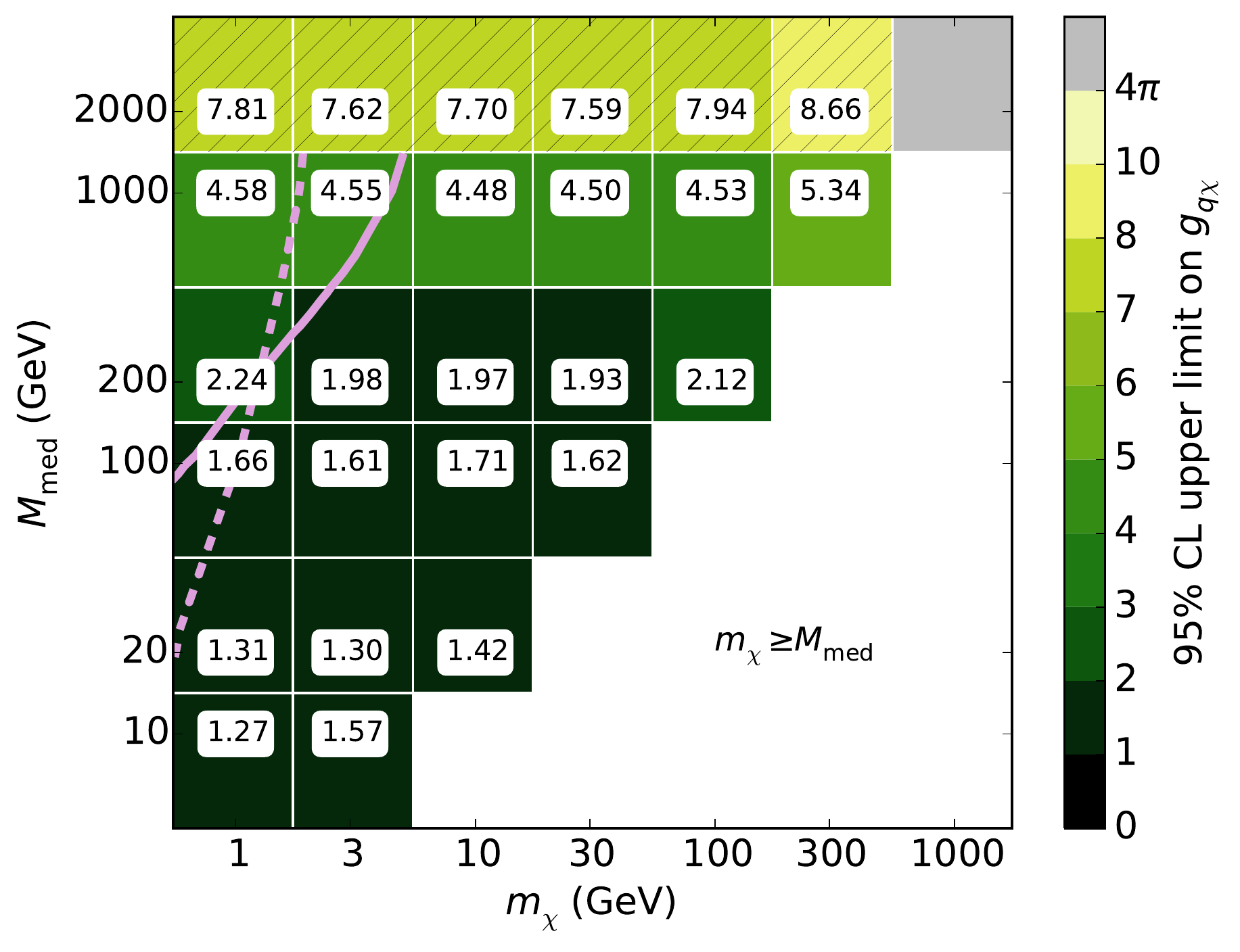}
    \caption{$tS$ model, mono-$Z$ channel.}
  \end{subfigure}
  \begin{subfigure}[t]{0.495\textwidth}
    \centering
    \includegraphics[width=1.\textwidth]{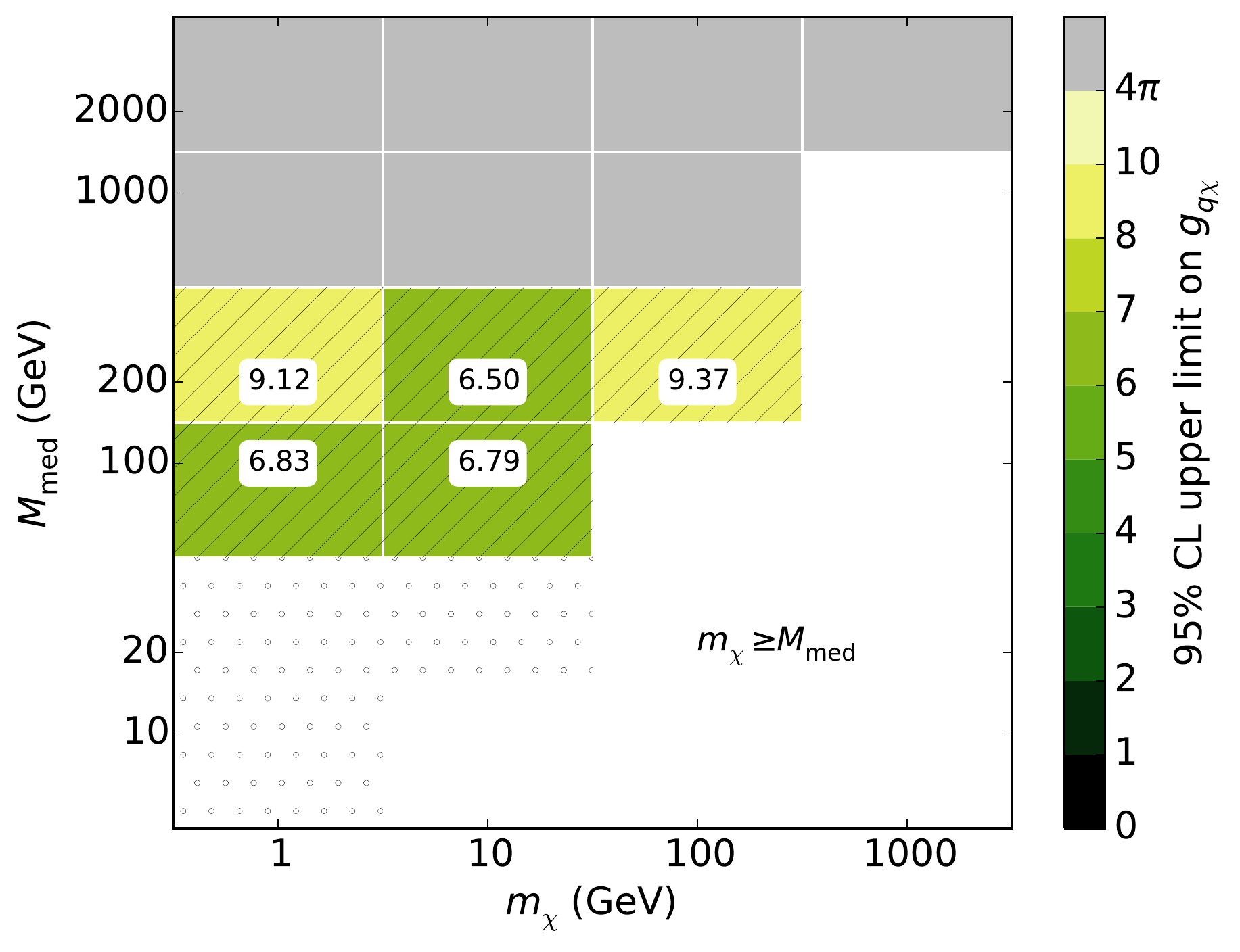}
    \caption{$tS$ model, mono-$W/Z$ channel.}
  \end{subfigure}
  \caption{Upper limits on the coupling $\gqX$ for the $t$-channel model in the \monoZ (left) and \monoWZ (right) channels. Refer to fig.~\ref{fig:results_sVsA_rat05} for details.}
  \label{fig:results_tS}
\end{figure}

In figs~\ref{fig:results_sVsA_rat05}-\ref{fig:results_tS} we also show the intercept line where constraints from direct detection experiments are equivalent to our \monoX constraints. Below and to the right of the dashed purple line, direct detection constraints are stronger while above and to the left of this line, the LHC gives the stronger limit. As with the relic density contours, we do not show the intercept where it does not pass through sufficient valid data points. We use the toolset from ref.~\cite{DelNobile:2013sia} to convert the strongest available direct detection constraints, which are from the LUX 2013 dataset~\cite{Akerib:2013tjd}, onto constraints on our models.

Compared to direct detection, the \monoX collider limits perform relatively better for the $sA$ model than for the $sV$ model. This is because the axial-vector coupling leads to a suppressed scattering rate in direct detection experiments while collider searches are relatively insensitive to the difference between the vector and axial-vector couplings. In the non-relativistic limit, the $tS$ model leads to a mix of both suppressed and unsuppressed operators.

The direct detection constraints assume that the DM candidate under consideration contributes 100\% of the local DM density, while the \monoX constraints make no assumptions about either the local DM density or overall abundance. In this sense the \monoX limits remain useful even in those regions of phase space where they are not as strong as those from direct detection.

\afterpage{\clearpage}

%% file: Conclusion.tex
In this paper we have examined a set of three simplified dark matter models, extracting constraints from ATLAS Run I mono-$X$ plus missing energy searches featuring the associated production of a mono-jet, mono-$Z(\rightarrow$ leptons), or mono-$W/Z (\rightarrow$ hadrons). We explored a phase space where both the DM and mediator masses span $\mathcal{O}$(GeV) to $\mathcal{O}$(TeV), and considered ratios of $\gX / \gq$ of 0.2, 0.5, 1, 2 and 5 in the $s$-channel models.

Rather than setting limits in the $\Mmed - \mDM$ plane for a fixed value of the coupling strength, we instead constrained the coupling strength as a function of both $\Mmed$ and $\mDM$ in a 3D plane. Whilst this approach necessitates the introduction of some approximations, it also allows for a thorough examination of the interplay between the DM production cross-section and the free parameters of the models.

As expected, the \monojet channel is found to yield the strongest limits on vector and axial-vector SM and DM couplings to a vector mediator exchanged in the $s$-channel. This channel is also found to perform well for small values of $\gX$. The limits obtained in the \monoZ channel, in comparison, are generally weaker by a factor of a few, while the \monoWZ results are weaker again. This is partly due to our conservative estimations of the systematic uncertainties and partly due to limited statistics resulting from a harder $\met$ selection cut. The width effects associated with the $t$-channel exchange of an SU(2) doublet scalar mediator are observed to vanish in both the \monoZ and \monoWZ channels, greatly simplifying the analysis of this model and confirming these as straightforward and competitive channels for future collider DM detection.

Where the axial-vector model is not excluded by perturbative unitarity requirements, we find the coupling limits to be on par with those of the vector model within each analysis channel. Weaker limits are found for the $t$-channel model, a result of cross-section suppression not present in the $s$-channel models.

Finally, we compared our limits to constraints from relic density and direct detection; although each search is subject to a different set of assumptions, this demonstrates the complementarity and impressive reach of simplified models as a tool for the interpretation of collider DM searches. We eagerly await the improved constraints expected from Run II of the LHC.

%% file: Acknowledgements.tex
A.J.B. and M.F.M. were supported by the Australian Research Council. J.G.~was supported by SNF (grant 200020\_156083). We thank Karl Nordstr{\"o}m for discussions on the cross-section reweighting, Brian Petersen, Steven Schramm, Rebecca Leane, Nicole Bell and Elisabetta Barberio for further helpful discussions, and Sean Crosby for technical support.

%% file: AppendixA.tex
In this appendix we present a summary of the procedure employed to calculate the 95\% confidence level (CL) limits on the coupling parameter $\sqrtgqgX$, where this parameter can be replaced with $\gqX$ for the $tS$ model, and $\Mstar$ in the validation of the \monojet analysis.

\subsection{Nominal values}
For each SiM, the nominal limit is calculated by taking the model-independent upper limit on $\sigma \times \mathcal{A} \times \epsilon$ from each analysis, dividing by the value of $\mathcal{A} \times \epsilon$ coming from signal MC (which is taken as a single parameter for each point) to obtain the limiting cross section $\sigma_{\mathrm{lim}}$, and rearranging eq.~\ref{eq:sigma_propto_couplings_schan} to convert to a limit on the couplings. In the $s$-channel on-shell case, the width can be expressed as a function of $\gq$ and the ratio $\gX / \gq$, which simplifies the calculation. We arrive at

\begin{equation}
  \sqrtgqgX_{\mathrm{lim}} =
  \begin{cases}
      \sqrtgqgX_{\mathrm{gen}} \times \left( \sigma_{\mathrm{lim}} / \sigma_{\mathrm{gen}} \right)^{\frac{1}{2}} & \mathrm{ if } \, \Mmed \geq 2 \mDM \,\, (s\mathrm{-channel})\\
      \sqrtgqgX_{\mathrm{gen}} \times \left( \sigma_{\mathrm{lim}} / \sigma_{\mathrm{gen}} \right)^{\frac{1}{4}} & \mathrm{ if } \, \Mmed < 2 \mDM \\
  \end{cases}
  \label{eq:nominal_limit}
\end{equation}
where $\sqrtgqgX_{\mathrm{gen}}$ and $\sigma_{\mathrm{gen}}$ are the input couplings and cross-section (taken from \PYTHIAnospace), respectively.

The signal region in each case is chosen based on where the best `expected' limit lies, where that limit is calculated assuming that exactly the expected SM background is observed.

\subsection{Uncertainty estimation}
\label{uncertainty_estimation_proc}
Our nominal limits rely on both $\sigma_{gen}$ and $\mathcal{A}\times\epsilon$ and so are subject to systematic uncertainties which derive from our choice of signal generation procedure. For our signal samples, there are three key sources of systematic uncertainty: the factorisation and renormalisation scales, the strong coupling constant ($\alpha_{s}$) and the choice of parton distribution function (PDF).

We assess the impact of the factorisation and renormalisation default scales in a straightforward manner; by varying them simultaneously by factors of 2 (`up') and 0.5 (`down'). The systematic effects of the strong coupling constant and PDF are difficult to separate and so are treated in tandem. We assume that the systematic uncertainty introduced by $\alpha_{s}$ at matrix-element level is negligible when compared to the PDF uncertainties, as demonstrated to be valid in ref. \cite{CERN-THESIS-2015-038}. The variation of $\alpha_{s}$ in conjunction with a change of PDF is done with the use of specific tunes in \PYTHIAnospace, which we change simultaneously with the PDF choice to estimate the uncertainty on $\Delta \sigma_{gen}$. The nominal choices of PDF and tune are varied `up' to NNPDF2.1LO PDF + Monash tune, and `down' to CTEQ6L1 PDF and ATLAS UE AU2-CTEQ6L1 tune. For the \monojet channel, the impact of the matching scale (QCUT) is assessed in a manner similar to that of the factorisation and renormalisation scales. That is, we vary the QCUT by factors of 2 (`up' to 160 GeV) and 0.5 (`down' to 40 GeV). These systematic uncertainty sources are summarised in table~\ref{tab:syst_unc}.

\begin{table}[!h]
\centering
\begin{tabular}{c|c|c|c}
\hline
\hline
main systematic & \multirow{2}{*}{PDF/tune} & factorisation and & matching scale \T \\
sources & & renormalisation scales & (\monojet only) \B \\
\hline
\multirow{2}{*}{variation `up'} & NNPDF2.1LO + & \multirow{2}{*}{2} & \multirow{2}{*}{160 GeV} \T \\
& Monash tune & & \B \\
& & & \\
\multirow{3}{*}{nominal} & MSTW2008lo68cl + & \multirow{2}{*}{1} & \multirow{2}{*}{80 GeV} \T \\
& ATLAS UE & & \B \\
& AU2-MSTW2008LO & & \B \\
& & & \\
\multirow{2}{*}{variation `down'} & CTEQ6L1 + & \multirow{2}{*}{0.5} & \multirow{2}{*}{40 GeV} \T \\
& ATLAS UE & & \B \\
& AU2-CTEQ6L1 & & \B \\
\hline
\hline
\end{tabular}
\caption{Reading left to right, the sources of systematic uncertainty considered in this analysis. Each point in phase space is varied up or down by one of these sources, and the systematic uncertainty is then taken from the resultant changes to the acceptance and cross-section in comparison to their nominal values.}
\label{tab:syst_unc}
\end{table}

The average variation in the nominal value of $\sigma_{\mathrm{lim}}$ (measured as a fraction of $\sigma_{\mathrm{lim}}$) resulting from each systematic source is added in quadrature and propagated to $\sqrtgqgX$ to obtain the total systematic uncertainty. This process is adjusted slightly to account for the inclusion of statistical uncertainties, which are estimated conservatively by taking the 95\% CL \emph{lower} limit on $\mathcal{A} \times \epsilon$ as calculated with the Wald approximation, i.e. $\mathcal{A}\times\epsilon \rightarrow (\mathcal{A}\times\epsilon) - \Delta(\mathcal{A}\times\epsilon)$. Note that the uncertainty on the luminosity is less than 3\%, so is considered to be negligible in comparison to other systematic sources.

%% file: AppendixB.tex
\subsection{Mono-jet channel}
\label{monojet_validation}
The signal generation and selection procedures for the \monojet channel are validated via reproduction of the ATLAS limits on $\Mstar \equiv \Mmed / \sqrtgqgX$, for the $s$-channel vector SiM. A comparison of SR7\footnote{We use this signal region as it is the only one for which ATLAS limits are provided.} limits for a representative sample of mediator masses with $\mX = $ 50 GeV, $\Gamma = M/8\pi$ and $\sqrtgqgX = 1$ is presented in table \ref{M_star_limits_monojet}. In general, good agreement is observed between the ATLAS and reproduced limits, with a maximum difference of 12\%. We note that a discrepancy of a few percent is expected given the differences in signal simulation. For example, the simplified matching procedure discussed in detail in Sec~\ref{matching_procedure} introduces an additional uncertainty of approximately 25\% for events with $\met > 350$ GeV when compared to the approach utilised by the ATLAS \monojet group. Further uncertainties are introduced by the jet smearing approximation used in place of a full detector simulation and by the 95\% CL estimation procedure (outlined in app.~\ref{Appendix_limitsetting}) used instead of a thorough HistFitter treatment. As our results are consistently more conservative than those of the ATLAS analysis, we consider our approach to be acceptable.

\begin{table}[!htbp]
\centering
\begin{tabular}{c|c|c|c}
 \hline
 \hline
 $\Mstar^{\tiny gen}$ & $\Mstar^{95\%\mathrm{CL}}$ [GeV] & $\Mstar^{95\%\mathrm{CL}}$ [GeV] & Difference \\
 $[$TeV$]$ & (ATLAS) & (this work) & $[\%]$ \\
 \hline
0.05 & 91 & 89 & 2.16 \\
0.3 & 1151 & 1041 & 7.3 \\
0.6 & 1868 & 1535 & 11.8 \\
1 & 2225 & 1732 & 12.0 \\
3 & 1349 & 1072 & 6.8 \\
6 & 945 & 769 & 8.5 \\
10 & 928 & 724 & 10.6 \\
30 & 914 & 722 & 9.6 \\
 \hline
 \hline
\end{tabular}
\caption{Comparison of the 95\% CL lower limits on $\Mstar$ from this work and from the ATLAS \monojet analysis \cite{Aad:2015zva}. The limits are for an $s$-channel vector mediator model with $\mX = $ 50 GeV and $\Gamma = \Mmed/8\pi$, and for the process $pp \rightarrow \chi \bar{\chi} + 1, 2j$ with QCUT = 80 GeV. Note that $\Mstar^{\tiny gen}$ is the input suppression scale.}
\label{M_star_limits_monojet}
\end{table}

\subsection{Mono-$Z$(lep) channel}
\label{monoZ_validation}

The ATLAS \monoZ results include an upper limit on the coupling $\gqX$ for a $t$-channel SiM analogous to our $tS$ model, and so it is this model which we use to validate our signal generation and selection procedures. Note that the following differences exist: the ATLAS model includes just two mediators ($up$- and $down$-type) where we consider six, the DM particle is taken to be Majorana where we assume Dirac, and the couplings $g_{t,b \chi}$ are set to zero where we have universal coupling to all three quark generations.

\begin{table}
\begin{center}
\begin{tabular}{ c | c | c | c | c }
\hline
\hline
$\mX$ & $\Mmed$ & $\gqX^{95\%\mathrm{CL}}$ & $\gqX^{95\%\mathrm{CL}}$ & Difference \T \\
$[$GeV$]$ & $[$GeV$]$ & (ATLAS) & (this work) & $[\%]$ \B \\
\hline
10 & 200 & 1.9 & 2.0 & 5.3 \T \\
 & 500 & 2.8 & 3.2 & 14.3 \\
 & 700 & 3.5 & 4.4 & 25.7 \\
 & 1000 & 4.5 & 5.2 & 15.6 \\
200 & 500 & 3.4 & 4.0 & 17.6 \T \\
 & 700 & 4.2 & 4.5 & 7.1 \\
 & 1000 & 5.2 & 5.3 & 1.9 \\
400 & 500 & 5.5 & 5.7 & 3.6 \T \\
 & 700 & 6.1 & 6.5 & 6.6 \\
 & 1000 & 7.2 & 7.4 & 2.8 \\
1000 & 1200 & 23.3 & 24.1 & 3.4 \T \B \\
\hline
\hline
\end{tabular}
\end{center}
\caption{Comparison of the 95\% CL upper limit on $\gqX$ from this work and from the ATLAS \monoZ analysis \cite{Aad:2014monoZlep}. The limits are for a variant of the $t$-channel scalar mediator model with Majorana DM for the process $pp \rightarrow \chi \bar{\chi} + Z (\rightarrow e^{+}e^{-}/\mu^{+}\mu^{-})$.}
\label{tab:monoZvalidation}
\end{table}

Table \ref{tab:monoZvalidation} shows the 95\% CL upper limits on $\gqX$ that we calculate using our own generation procedure (and the values in table~\ref{tab:sigmalim_monoZ}), compared with the limits taken from the ATLAS analysis. Also shown is the difference as a percentage of the ATLAS limit. We see reasonable agreement; most of the 11 points in parameter space are within 10\% of the ATLAS limits, and all are within 26\%. Additionally, our results are consistently more conservative, which again is to be expected given the less sophisticated nature of our generation procedure. As in the case of the \monojet validation, the differences stem from the use of $p_{\mathrm{T}}$ smearing applied to the leptons (rather than a full reconstruction simulation) and from the simplified treatment of systematics; we also obtained $\sigma \times \mathcal{A} \times \epsilon$ independently using the public results.

\subsection{Mono-$W/Z$(had) channel}
\label{monoWZ_validation}

The event generation and selection procedures for the \monoWZ channel are validated via reproduction of the ATLAS limits on $\Mstar$ for the D5 and D9 effective operators with $\mX = $ 1 GeV, using the upper limits on $\sigma \times \mathcal{A} \times \epsilon$ listed in table~\ref{tab:sigmalim_monoWZ}. We see agreement within 12.5\% and 7.4\% respectively, where the ATLAS limits are consistently stronger, as shown in table \ref{tab:monoWZvalidation}. The relative sizes of the discrepancies are expected given that only low-$\met$ limits are available for the D5 operator while we use the high-$\met$ signal region in our recast. Note that a general discrepancy of a few percent is expected for both operators for the reasons discussed in sections \ref{monojet_validation} and \ref{monoZ_validation}, and also because we use a cut-and-count approach while the ATLAS limits are extracted using a shape-fit. Furthermore, the ATLAS limits are quoted at 90\% CL while ours are calculated at 95\% CL.

\begin{table}[!h]
\begin{center}
\begin{tabular}{ c | c | c | c | c }
\hline
\hline
EFT operator & $\mX$ & $\Mstar^{90\%\mathrm{CL}}$ $[$GeV$]$ & $\Mstar^{95\%\mathrm{CL}}$ $[$GeV$]$  & Difference \T \\
&$[$GeV$]$ & (ATLAS) & (this work) & $[\%]$ \B \\
\hline
D9 & 1 & 2400 & 2221 & 7.4 \\
D5 & 1 & 570 & 499 & 12.5 \\
\hline
\hline
\end{tabular}
\end{center}
\caption{Comparison of the 95\% CL lower limits on $\Mstar$ from this work and from the ATLAS \monoWZ analysis \cite{Aad:2013monoWZ}. The limits correspond to the process $pp \rightarrow \chi \bar{\chi} + W/Z$ ($\rightarrow jj$).}
\label{tab:monoWZvalidation}
\end{table}